\documentclass[]{aa}

\usepackage{natbib}
\usepackage{breqn}
\usepackage{xcolor}
\usepackage{graphicx}
\usepackage{media9}
\usepackage{txfonts}
\usepackage{multirow}
\usepackage{hyperref} 
\hypersetup{
    colorlinks = true,
    linkbordercolor = {cyan},
    linkcolor={blue},
    citecolor={blue},}

\begin{document}

\def\figureautorefname{Fig.}
\def\equationautorefname~#1\null{Eq. (#1)\null}

\def\sectionautorefname{Sect.}
\def\subsectionautorefname{Sect.}
\def\subsubsectionautorefname{Sect.}

\title{Resonance locking and tidal evolution in rotating $\gamma$-Doradus binaries}
\titlerunning{MAD tides}

\author{L. Fellay\inst{1} \and M.-A. Dupret\inst{1}  \and P. A. Ko{\l}aczek-Szyma{\'n}ski \inst{1,2}}
\institute{STAR Institute, University of Liège, 19C Allée du 6 Août, B$-$4000 Liège, Belgium \and Astronomical Institute, Faculty of Physics and Astronomy, University of Wroc\l aw, ul. Kopernika 11, 51-622 Wroc\l aw, Poland}
\date{January,  2024}

\abstract{In binary systems, studying tidal interactions is key to understanding the evolution of binary populations. From a theoretical standpoint, the primary dissipation process occurring in stars with radiative envelopes is believed to be radiative damping of high-radial-order tidally excited oscillations, which is in agreement with observations of most binary systems. However, recent studies have suggested that outside this dissipation regime, dynamical tides can act in the opposite manner (a phenomenon known as `inverse tides'), and resonance locking could significantly impact the orbital evolution of binary systems. While the basic theoretical principles behind inverse tides are understood, previous studies have primarily used simplified models that assume that resonance locking with a single oscillation mode occurs in order to assess its impact over long timescales without evaluating its feasibility or determining the orbital and rotational conditions under which locking should occur.}
{We aim to study inverse tides and resonance locking by simultaneously including the effect of all the forcing frequencies and accounting for the effect of the rotation on the forced oscillations.}
{We have developed an orbital evolution code that is coupled to a stellar oscillation code to compute on the fly the impact of dynamical tides on the rotational and orbital evolution of binary systems including multiple simultaneous forcing frequencies.}
{We find that resonance locking can be stable over a long period of time and a source of long-term exchange of angular momentum for rapidly rotating stars (with rotation greater than $15\%$ of the critical rotation). Long-term locking can increase the total angular momentum of a fast-rotating star by approximately 70$\%$ during the main sequence.  For slow-rotating stars, resonance locking can slow down the rotational evolution of the system over most of the main-sequence phase, even in the presence of strong tidal interactions. }
{Inverse tides and resonance locking are mechanisms capable of preserving a signature of the initial stellar rotation in binary systems, even when non-resonant tidal interactions are expected to be strong. This mechanism efficiently drives asynchronisation in binary systems where significant discrepancies already exist between the orbital and rotational frequencies.}
\keywords{Stars: oscillations- Binaries: general - Binaries: close - Stars: interiors - Stars: evolution - Celestial mechanics}
\maketitle

\noindent
\section{Introduction}
Unlike stars with convective envelopes, those with radiative envelopes dissipate most of their orbital energy through high-radial-order tidally excited oscillations. The excitation of oscillation modes by tides and their subsequent damping effects on the orbit is a process known as dynamical tides. In general, when the forcing frequency from the companion ($\sigma_{m,k}$) is small compared to the inverse of the stellar dynamical timescale ($1/\tau_{\mathrm{dyn}}$), the tidally excited oscillations are damped in the near-surface layers by radiative damping \citep{Zahn1975,Zahn1977}.  This regime, for which the theory of tidal dissipation formulated by \cite{Zahn1975,Zahn1977}  is valid,  predicts a dissipation of orbital energy into the star.  The dissipation ultimately drives the system towards a minimum energy state, which corresponds to a circular orbit with the stellar rotation rate synchronised to the orbital motion. If the system has not yet reached a circularisation, a pseudo-synchronisation can occur where the rotation and orbital angular velocity at the periastron are near synchronisation \citep{Hut1981}.  For higher forcing frequencies, tidally excited oscillations can resonate with lower-radial-order oscillation modes, induce non-linear mode excitations \citep{Weinberg2012,Guo2020,Guo2022}, or generate travelling waves \citep{Burkart2012,Bowman2019,Ratnasingam2019}. Each of these phenomenon results in a non-linear response of the companion to the forcing frequency \citep{Zahn1977,Witte1999a,Witte1999b,Witte2001,Savonije2002,Savonije2005}, making the classical theory of \cite{Zahn1975,Zahn1977} inapplicable.  Furthermore, \cite{Zahn1975,Zahn1977} neglected the effects of rotation on the mode eigenfunctions, which may not be valid for rapidly rotating stars.
\\~\\For high forcing frequencies relative to the inverse of the stellar dynamical timescale, \cite{Fuller2021} points out that in stars with unstable modes, dynamical tides can extract energy and angular momentum from the star and transfer them to the orbit. This phenomenon,  called `inverse tides', allows for rapid variations in stellar rotation (as the stellar moment of inertia is significantly lower than that of the orbit),  with a sign opposite to that predicted by \cite{Zahn1975,Zahn1977}.  Only stars within an instability strip can experience this process, as unstable modes must be excited by tides \citep{Fuller2021}.  In particular, intermediate-mass stars (from 1.5 $M_\odot$ to 8.0 $M_\odot$ in this article) in $\delta$-Scuti, $\gamma$-Doradus, and slowly pulsating B-type (SPB) stars, instability strips present an excellent opportunity to study inverse tides, given the extensive research on their structure, pulsations, and rotation \citep{Guzik2000,Dupret2002,Dupret2005,Townsend2005,Bouabid2013,Li2019,Li2020a,Li2020,Saio2021}.
\\~\\In the framework of inverse tides, the primary mechanism for long-term angular momentum exchange is resonance locking \citep{Fuller2017,Fuller2021}. This mechanism occurs when a binary system maintains a state in which the variation of a forcing frequency matches the change in resonance frequency of an unstable mode due to stellar evolution. In \cite{Fuller2021}, resonance locking is assumed to take place; however, the effect of multiple simultaneous oscillation modes and the effects of rotation or evolution on mode properties are neglected as well as the modification of the forcing frequencies consequent to the impact of tides. These factors could significantly influence the feasibility of resonance locking, as additional oscillation modes might disrupt the locking process, and rotation could alter mode properties and resonance frequencies \citep{Unno1989}.
\\~\\In general,  tides remain a challenging aspect of binary evolution modelling.  Observationally, constraints suggest that Zahn’s tidal theory agrees with the global trends of circularisation \citep{Khaliullin2010} and synchronisation (or pseudo-synchronisation) \citep{Khaliullin2007} in binary star populations, particularly among low-mass stars \citep{Shporer2016,Lurie2017}.  For intermediate-mass stars, the limited number of studies suggests that the circularisation and synchronisation trends (or pseudo-synchronisation) observed in close-in binaries are mostly consistent with Zahn’s theory \citep{Khaliullin2007,Khaliullin2010,Zimmerman2017,Saio2022}.  In contrast,  studies of open clusters often show significant discrepancies.  For instance,  reproducing the observed circularisation period in NGC 6819 with the standard prescription used in standard population synthesis codes such as BSE \citep{Hurley2002} requires artificially increasing the convective damping efficiency by a factor of 100 \citep{Meibom2005, Geller2013,Milliman2014}.
On the other hand,  the inverse tides mechanism predicts the existence of two distinct sub-populations. The first consists of unsynchronised rapidly rotating stars (or a population of Be stars formed without prior mass transfer) exhibiting tidally forced oscillations even in close binaries, a prediction that seems inconsistent with observations,  except for a few fast-rotating outliers, such as HD 9899216 \citep{Thompson2012,Saio2022}, $\psi$ Centauri, \citep{Mantegazza2010}, and KIC 4659476 \citep{Shporer2016}, for which synchronisation may not yet have occurred even with the formalism of \cite{Zahn1975,Zahn1977}. These exceptions suggest that inverse tides may not play a dominant role, necessitating further investigation. However, a second population of slowly rotating stars is also predicted by \cite{Fuller2021}. This prediction appears to be more consistent with observations, as more stars with sub-synchronous rotation are detected \citep{Zimmerman2017,Lampens2018,Guo2019,Li2020a,Li2020}, a feature that is not predicted in the framework of \cite{Zahn1975,Zahn1977} for close binary systems. In addition, \cite{Zanazzi2021} points out that resonance locking may be a prominent mechanism occurring during the stellar pre-main sequence of solar-like and intermediate-mass stars to circularise the orbit. However, in this article we focus on resonance locking during the main sequence.
\\~\\In this study, we conduct an in-depth analysis of resonance locking. We have developed a new computational tool, Tidal and Rotational Evolution by Modeling Oscillations and Resonances (TREMOR), capable of integrating the secular orbital evolution timescale due to dynamical tides.  By computing tidally excited oscillations on the fly using our linear non-adiabatic oscillation code MAD \citep{Dupret2001,Dupret2002,Bouabid2013,Fellay2025}, which includes the effects of rotation on mode eigenfunctions via the traditional approximation, we account for the contributions of all dynamically excited tides within the system. This approach allows us to accurately predict the feasibility of resonance locking and its long-term impact on the orbital and rotational evolution of binary systems.
\\~\\In \autoref{sect_TREMOR}, we formulate the problem of dynamical tides using the formalism of \cite{Smeyers1998,Willems2002,Willems2003,Willems2010,CAFeiN2013} and describe in detail the methodology implemented in TREMOR to solve these equations simultaneously for all system forcing frequencies.  \autoref{sect_intergration_dy} presents the results from TREMOR’s full numerical integrations for a few scenarios where dynamical tides play a key role.  Finally, we discuss our findings in and conclude in \autoref{sect_discussion}.

\section{Dynamical tides in fast-rotating binaries}\label{sect_TREMOR}
\subsection{Formulation of the problem }\label{subsect_formulation_dy_tides}
In this section, we present the mathematical formalism used to describe the forcing of oscillation modes in a star and its consequent impact on the rotational and orbital properties of a binary system following the approach of \citet{Smeyers1998,Willems2002,Willems2003,Willems2010,CAFeiN2013}. We adopted a direct modelling approach \citep{Savonije1983,Savonije1984,Smeyers1998,Willems2003,Burkart2012,CAFeiN2013,gyre_tides2023} where a stellar oscillation code or a non-adiabatic oscillation code, MAD \citep{Dupret2001,Dupret2002,Bouabid2013,Fellay2025} in our case, is modified to include the forcing from the companion (this forced non-adiabatic oscillation code is called MAD_tides).  In MAD_tides, we included the time-dependent convection model of  \cite{Grigahcene2005} for our non-adiabatic computations in $\gamma-$Doradus stars.
In our modelling, we assumed a solid body rotation and included the effect of rotation in our oscillation code through the traditional approximation without employing the Cowling approximation. (For further details on our implementation of rotation, see \cite{Fellay2025}.) In this work, we adopted the `uncoupled approximation' formalism presented in \cite{Fellay2025}. The amplitude of the tidally excited oscillations induced in a star by its companion is proportional to the tidal potential perturbation from the companion.
\\~\\The gravitational potential from the companion is obtained classically using the perturbative approach, which assumes that stellar deformations are small perturbations from the spherical symmetry compared to the radius of the star. In this work, as is classically done, the secondary was approximated as a point mass. Within this approximation and in the perturbative framework, the tidal perturbing potential from the companion can be expressed as a Fourier series in a coordinate frame centred on the primary:
\begin{align}\label{eq_pot_surf_bcond}
\psi_2(r, \mu, \phi, \mathcal{M})&=-\dfrac{GM_2}{r}\sum _{\ell\geq 2} \sum _{m=-\ell} ^{m=\ell} \sum _{k=-\infty} ^{k=\infty} \left(\dfrac{r}{a} \right)^{\ell+1} \\ & X^{-(\ell+1),-m}_{k}e^{i (k \mathcal{M}+m\phi)} d_{\ell}^{m} P_{\ell}^{m}(\mu), \nonumber
\end{align}
 where $\mathcal{M}$ is the mean anomaly ($\mathcal{M} = \Omega_{\mathrm{orb}} t$, with $\Omega_{\mathrm{orb}}$ being the orbital angular frequency), $G$ is the gravitational constant, $a$ is the semi-major axis, $M_2$ is the mass of the companion, $r$ is the radial coordinate of the primary star, $\mu$ is the cosinus of the co-latitude angle, $\phi$ is the azimuthal angle, $\ell$ and $m$ are respectively the spherical degree and azimutal order of the associated Legendre polynomials $P_{\ell}^{m}(\mu)$, $k$ are the orbital harmonics, and $d_{\ell}^{m}$ is a normalisation coefficient defined as
 \begin{equation}
d^m_{\ell}= \dfrac{(\ell-m) !}{(\ell+m )!}P_{\ell}^{m}(0).
\end{equation}  
The coefficients $X^{-(\ell+1),-m}_{-k}$ are the Hansen coefficients defined by
\begin{equation}
X^{n,-m}_{k}=\dfrac{(1-e^2)^{n+3/2}}{2\pi}\int_{0}^{2\pi} \dfrac{\cos{(m v+k \mathcal{M})}}{(1+e\cos v)^{n+2}} \mathrm{d}v,  \nonumber
\end{equation}
where $v$ is the true anomaly and $e$ is the orbital eccentricity.
Within our formalism, a modified dimensionless radial potential perturbation from the companion, used as the inhomogeneous term in the forced oscillation problem (see \cite{Savonije1983,Savonije1984,Smeyers1998,Willems2003,Burkart2012,CAFeiN2013,gyre_tides2023,Fellay2025}), is defined as
\begin{equation}\label{eq_tidal_perturbation}
\psi_{2,\ell,k}^m(r)=-\left(\dfrac{r}{R_1} \right)^{\ell},
\end{equation}
where $R_1$ is the radius of the primary star. Because this term is independent of the orbital parameters, the dimensionless measure of the star's response, $\mathrm{Im}(F_{\ell,m,k})$, to the various forcing angular frequencies defined by \cite{Smeyers1998,Willems2003,Willems2010,CAFeiN2013} is given by
\begin{equation}\label{eq_norm_Im_Flmk}
\mathrm{Im}(F_{\ell,m,k}) = -\dfrac{1}{2}\mathrm{Im} (\psi^{\prime}_{\ell,m,k}(R_1)),
\end{equation}
where $\psi^{\prime}_{\ell,m,k}(R_1)$ is the dimensionless Eulerian potential perturbation response evaluated at the stellar surface by the non-adiabatic oscillation code. All the physics of the oscillations, as well as their influence on the orbital evolution of the system, is contained in $\psi^{\prime}_{\ell,m,k}(R_1)$. The Fourier coefficients $c_{\ell,m,k}$ from \cite{Smeyers1998,Willems2003,Willems2010,CAFeiN2013} are expressed as
\begin{equation}
c_{\ell,m,k}(e)= X^{-(\ell+1),-m}_{k} d_{\ell}^{m}\left( \dfrac{R_1}{a}\right)^{\ell-2}.
\end{equation}
Using an inhomogeneous term that is independent of the orbital configuration is particularly advantageous, as it allows the stellar response to be computed separately from the system’s orbital properties and later rescaled to match the specific parameters of the binary system.
In the case of circular orbits, we verified our numerical computation of $c_{\ell,m,k}(e)$ with the values provided by \citet{Willems2010}. The effect of the excitation of oscillation modes by the forcing of the companion on the orbital parameters of the system is then given by
\begin{equation}\label{eq_secular_a}
\dfrac{\mathrm{d}a}{\mathrm{d}t} =4\Omega_{\mathrm{orb}}\dfrac{M_2}{M_1}a\sum_{\ell=2}^{\infty} \sum_{m=-\ell}^{\ell} \sum_{k=-\infty}^{\infty} \left( \dfrac{R_1}{a} \right)^{\ell+3} \mathrm{Im}(F_{\ell,m,k}) G^{(2)}_{\ell,m,k}(e),
\end{equation}
\begin{equation}\label{eq_secular_e}
\dfrac{\mathrm{d}e}{\mathrm{d}t} =4\Omega_{\mathrm{orb}}\dfrac{M_2}{M_1}\sum_{\ell=2}^{\infty} \sum_{m=-\ell}^{\ell} \sum_{k=-\infty}^{\infty} \left( \dfrac{R_1}{a} \right)^{\ell+3} \mathrm{Im}(F_{\ell,m,k}) G^{(3)}_{\ell,m,k}(e),
\end{equation}
\begin{align}\label{eq_secular_rot}
&\dfrac{1}{I}\dfrac{\mathrm{d}L_{\mathrm{rot}}}{\mathrm{d}t}=\dfrac{\mathrm{d}\Omega_{\mathrm{rot}}}{\mathrm{d}t}+\Omega_{\mathrm{rot}}\dfrac{\mathrm{d}\ln I}{\mathrm{d}t}= \\ &4\dfrac{\Omega_{\mathrm{orb}}}{I}\left(\dfrac{G a}{M_1+M_2}\right)^{1/2} M_2^2\sum_{\ell=2}^{\infty} \sum_{m=-\ell}^{\ell} \sum_{k=-\infty}^{\infty}  \left( \dfrac{R_1}{a} \right)^{\ell+3} \mathrm{Im}(F_{\ell,m,k}) G^{(4)}_{\ell,m,k}(e),\nonumber 
\end{align}
where $M_1$ is the mass of the primary, $L_{\mathrm{rot}}$ is the rotational angular momentum of the primary, $I$ is its moment of inertia,  and  $\Omega_{\mathrm{rot}}$ is its rotational angular frequency. The evolution of the stellar rotation rate can be decomposed into a tidal component $\left(\mathrm{d}\Omega_{\mathrm{rot}}/\mathrm{d} t\right)_{\mathrm{tides}}$ and an evolutionary component $\left(\mathrm{d}\Omega_{\mathrm{rot}}/\mathrm{d} t\right)_{\mathrm{evo}}$:
\begin{equation}\label{eq_decomposition_rotation}
\dfrac{\mathrm{d}\Omega_{\mathrm{rot}}}{\mathrm{d} t}=\left(\dfrac{\mathrm{d}\Omega_{\mathrm{rot}}}{\mathrm{d} t}\right)_{\mathrm{tides}}+\left(\dfrac{\mathrm{d}\Omega_{\mathrm{rot}}}{\mathrm{d} t}\right)_{\mathrm{evo}},
\end{equation}
where the evolutionary component is given by the right-hand side of the second equality in \autoref{eq_secular_rot}, 
\begin{equation}
\left(\dfrac{\mathrm{d}\Omega_{\mathrm{rot}}}{\mathrm{d} t}\right)_{\mathrm{evo}}=-\dfrac{\mathrm{d} \ln I_\star}{\mathrm{d} t}\Omega_{\mathrm{rot}},
\end{equation} 
and the tidal component is given by the third equality of \autoref{eq_secular_rot}.
The coefficients $G^{(2)}_{\ell,m,k}(e)$,$G^{(3)}_{\ell,m,k}(e)$, and $G^{(4)}_{\ell,m,k}(e)$ are defined as
\begin{align}
G^{(2)}_{\ell,m,k}&(e)=\dfrac{2}{\pi(1-e^2)^{\ell+1}} c_{\ell,m,k}(e) P_{\ell}^{\vert m \vert}(0)\\ &\times \left[(\ell+1)e \int^{\pi}_0 (1+e \cos v)^{\ell} \sin(m v+k \mathcal{M}) \sin v\mathrm{d} v \right. \nonumber \\ &- \left.  m \int^{\pi}_0 (1+e \cos v)^{\ell+1} \cos(m v+k \mathcal{M}) \mathrm{d} v  \right],\nonumber 
\end{align}
\begin{align}
&G^{(3)}_{\ell,m,k}(e)=\dfrac{2}{\pi(1-e^2)^{\ell+1}} c_{\ell,m,k}(e) P_{\ell}^{\vert m \vert}(0)\\ &\times \left[(\ell+1)e \int^{\pi}_0 (1+e \cos v)^{\ell} \sin(m v+k \mathcal{M}) \sin v\mathrm{d} v \right. \nonumber \\ &- \left.  m \int^{\pi}_0 (1+e \cos v)^{\ell-1} \left[(1+e\cos v)^2 -(1-e^2)\right] \right. \nonumber  \\ & \times \left.  \dfrac{}{}\cos(m v+k \mathcal{M}) \mathrm{d} v  \right],\nonumber 
\end{align}
\begin{equation}
G^{(4)}_{\ell,m,k}(e)=\dfrac{e}{(1-e^2)^{1/2}}\left[G^{(3)}_{\ell,m,k} -\dfrac{1-e^2}{2e}G^{(2)}_{\ell,m,k}  \right].
\end{equation}
Our computations of  $G^{(2)}_{\ell,m,k}(e)$,$G^{(3)}_{\ell,m,k}(e)$, and $G^{(4)}_{\ell,m,k}(e)$  were verified with the values given in \cite{Willems2010} for circular orbits. Finally, for convenience, we introduced the characteristic timescales of orbital and rotational evolution, which were used to determine the time taken by tides to drive migration ($\tau_a$), circularise the orbit ($\tau_e$), modify the stellar rotation ($\tau_{\Omega_{\mathrm{rot}}}$), and alter the angular momentum ($\tau_{L_{\mathrm{rot}}}$):
\begin{equation}\label{eq_characteristic_timescales}
\tau_a=\dfrac{a}{\mathrm{d}a/\mathrm{d} t};\ \tau_e=\dfrac{e}{\mathrm{d}e/\mathrm{d} t}; \  \tau_{\Omega_{\mathrm{rot}}}=\dfrac{\Omega_{\mathrm{rot}}}{\mathrm{d}\Omega_{\mathrm{rot}}/\mathrm{d} t}; \  \tau_{L_{\mathrm{rot}}}=\dfrac{L_{\mathrm{rot}}}{\mathrm{d}L_{\mathrm{rot}}/\mathrm{d} t}.
\end{equation}
\subsection{Principle behind resonance locking}\label{subsect_principle_locking}
\begin{figure}[h]
\centering
\includegraphics[width=\hsize]{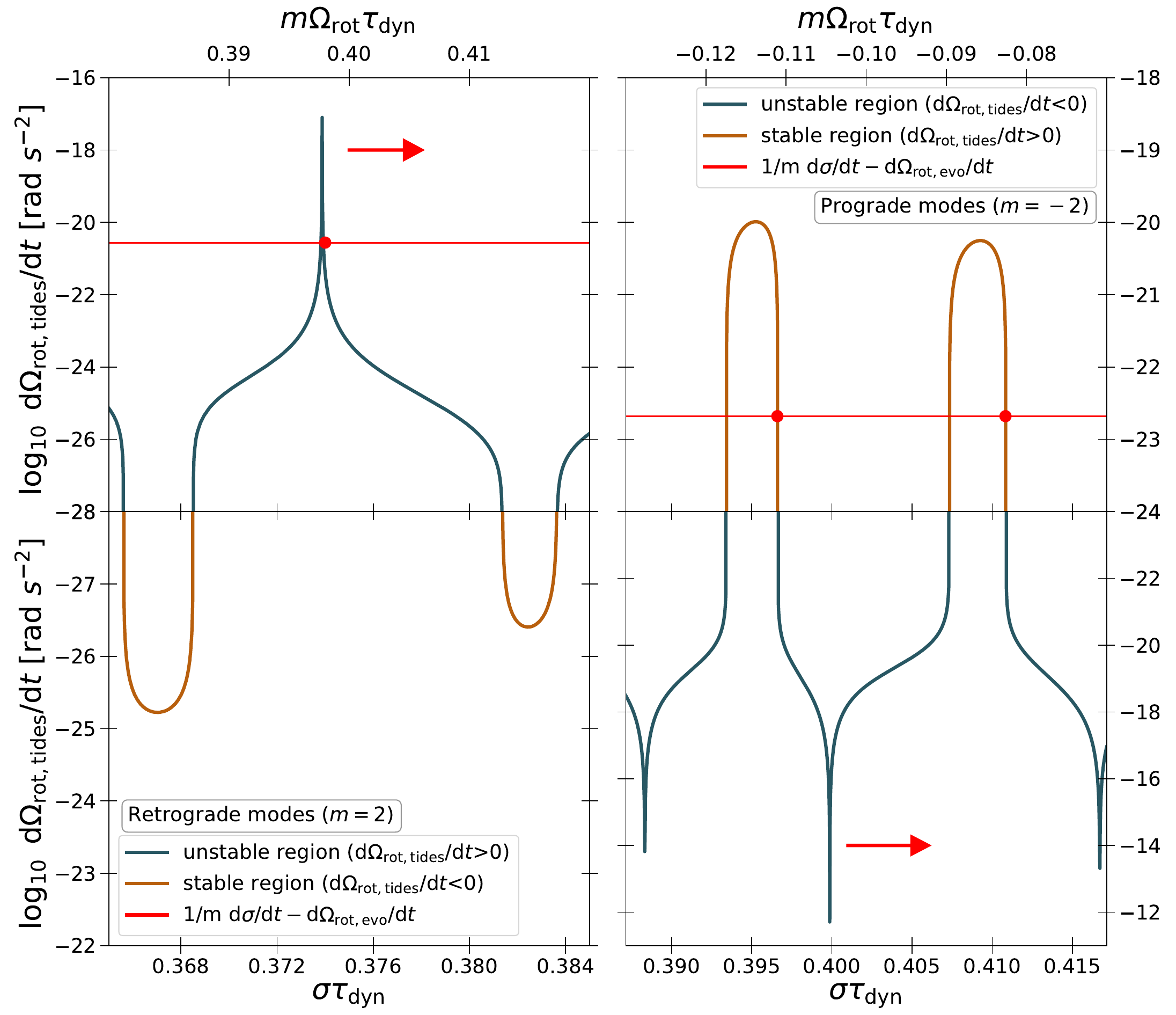}
\caption{Comparison of $1/m\ \mathrm{d}\sigma/\mathrm{d}t - \left(\mathrm{d}\Omega_{\mathrm{rot}}/\mathrm{d}t\right)_{\mathrm{evo}}$ (red curve) with $\left(\mathrm{d}\Omega_{\mathrm{rot}}/\mathrm{d}t\right)_{\mathrm{tides}}$ for a retrograde mode ($\ell=2$, $m=2$, $k=-2$, $n=30$, $\Omega{\mathrm{rot}}=0.2\  \Omega_{\mathrm{crit}}$; left) and a prograde mode ($\ell=2$, $m=-2$, $k=2$, $n=20$, $\Omega_{\mathrm{rot}}=0.05\ \Omega_{\mathrm{crit}}$; right). Forcing frequencies have been normalised by the dynamical timescale, $\tau_{\mathrm{dyn}}$. Blue segments represent unstable mode regions; orange curves represent stable regions. Arrows show the direction of mode evolution and spin trends. Red dots mark resonance locking points. Animated versions of this figure illustrating the various possible initial configurations and outcomes are provided in \autoref{apdx_animation_locking}.}
\label{fig.illustration_resonance_locking}
\end{figure}

One of the key phenomenon induced by dynamical tides that has to be reproduced by TREMOR is resonance locking. This mechanism has been introduced in several works (see e.g. \cite{Fuller2017,Fuller2021}). The primary contribution of this study is to provide a quantitative verification of the feasibility of resonance locking while accounting for the effects of rotation, the evolution of forcing frequencies,  stellar evolution and interactions between different forced oscillation modes.
Resonance locking occurs when the co-rotating-frame forcing frequency associated with a given $(m,k)$ pair (see \autoref{eq_forcing_frequency}) evolves at the same rate as the co-rotating frequency of the closest free oscillation mode with the radial order $n$.  In such cases, the de-tuning, defined as the difference between the forcing and co-rotating frequency, remains nearly constant. More specifically, resonance locking occurs when the resonance exerts a repulsive effect that prevents the de-tuning from decreasing.  In this regime, a smaller de-tuning leads to a stronger tidal response, which in turn pushes the forcing frequency towards a larger de-tuning. If the star's structural evolution tends to push the system away from resonance (e.g. through envelope expansion and associated spin-down), resonance locking can maintain the system in resonance.
This mechanism can remain stable over stellar evolution timescales, as we later show, and dynamical tides provide a long-term source of angular momentum and energy exchange between the components of the system and their orbit. However, not all oscillation modes are capable of supporting resonance locking. In this section, we present the basic principles of resonance locking and identify the types of modes that can support it.
\\For stars within the $\gamma$-Doradus or SPB instability strips, the co-rotation frequencies of gravito-inertial modes generally increase over time due to the rising Brunt–Väisälä frequency in the radiative zone near the contracting core. For a given orbital angular frequency and stellar rotation frequency, the forcing frequency is given by
\begin{equation}\label{eq_forcing_frequency}
\sigma_{m,k}= k\Omega_{\mathrm{orb}}  +m\Omega_{\mathrm{rot}}.
\end{equation}
where we adopt the sign convention for $k$ from \cite{CAFeiN2013}, which differs from that used in \cite{gyre_tides2023}. We ensured $\sigma_{m,k} > 0$ throughout. If a negative frequency is obtained, the same oscillation mode can be studied by changing the signs of both $m$ and $k$. Under this convention, $k m < 0$ corresponds to prograde waves in the inertial frame, and $k m > 0$ corresponds to retrograde waves. Prograde inertial-frame waves tend to have higher amplitudes due to larger Hansen coefficients. Thus, in this work, we consider only prograde waves in the inertial frame. Additionally,  in the co-rotating frame, $m > 0$ corresponds to retrograde waves, and $m < 0$ corresponds to prograde waves. Unless explicitly stated, references to `prograde' or `retrograde' denote the motion in the co-rotating frame, which is most relevant to the physics of modes.
\\For $\ell=2$ prograde modes ($m = -2$, $k > 0$), the forcing frequency is always positive when $2 \Omega_{\mathrm{rot}} < \Omega_{\mathrm{orb}}$. In this case, resonances with stable prograde modes mimic the behaviour of equilibrium tides, where angular momentum is transferred from the orbit to the star. This leads to spin-up of the star and orbital decay, driving the system towards synchronisation.  On the contrary, resonances with unstable prograde modes result in angular momentum transfer from the star to the orbit, spinning down the star and increasing the semi-major axis. 
\\During resonance locking, the forcing frequency $\sigma_{m,k}$ must evolve at the same rate as the co-rotating frequency $\sigma$ of the closest free oscillation mode. Neglecting changes in $\Omega_{\mathrm{orb}}$ (which is justified because the orbital moment of inertia is much greater than that of the star), the locking condition becomes
\begin{equation}
\dfrac{\mathrm{d}\sigma}{\mathrm{d}t}= m\left(\dfrac{\mathrm{d}\Omega_{\mathrm{rot}}}{\mathrm{d} t}\right)_{\mathrm{tides}}+m\left(\dfrac{\mathrm{d}\Omega_{\mathrm{rot}}}{\mathrm{d} t}\right)_{\mathrm{evo}}.
\end{equation}
The locking condition can finally be rewritten to isolate the tidal contribution as
\begin{equation}\label{eq_basic_resonance_criter}
\left(\dfrac{\mathrm{d}\Omega_{\mathrm{rot}}}{\mathrm{d} t}\right)_{\mathrm{tides}}=\dfrac{1}{m}\dfrac{\mathrm{d}\sigma}{\mathrm{d}t}-\left(\dfrac{\mathrm{d}\Omega_{\mathrm{rot}}}{\mathrm{d} t}\right)_{\mathrm{evo}}.
\end{equation}
In \autoref{fig.illustration_resonance_locking}, we illustrate resonance locking for the prograde and retrograde modes in a twin 1.5 M$_\odot$ $\gamma$-Doradus binary system at an age of 738 Myr (see \autoref{subsect_reference_model} for model details). The red curves show the right-hand side of \autoref{eq_basic_resonance_criter}, while the tidal contribution $\left( \mathrm{d} \Omega_{\mathrm{rot}} / \mathrm{d}t \right)_{\mathrm{tides}}$ as obtained with MAD_tides and computed via \autoref{eq_secular_rot} is plotted for comparison.
In practice, not all resonance lockings are stable. In the left panel of \autoref{fig.illustration_resonance_locking}, no stable locking is found to the left of the peak, even though the tidal amplitude reaches its maximum. In this region, if the system lies below the red threshold, the mode drifts away from the potential locking position.  On the contrary,  if the system lies above the red threshold, tides modify the forcing frequency more rapidly than the intrinsic mode evolution, and the system is driven towards the locking position on the right. An analogous reasoning applies to the right panel. For the case of retrograde modes, we provide in the supplementary material of this article,  animations illustrating the locking process from different initial positions around the resonance.
For retrograde modes, when $\left(\mathrm{d}\Omega_{\mathrm{rot}}/\mathrm{d} t\right)_{\mathrm{tides}}<1/m\ \mathrm{d}\sigma/\mathrm{d}t-\left(\mathrm{d}\Omega_{\mathrm{rot}}/\mathrm{d} t\right)_{\mathrm{evo}}$, the motion of the modes resonances dominates the variation of the de-tuning until a locking is reached and the de-tuning remains constant. Resonance locking typically occurs in unstable regions near strong resonances. This leads to strong angular momentum deposition into the star and rotational spin-up.
In contrast, prograde modes undergo resonance locking in stable low-amplitude regions of the spectrum. However, such locking is only made possible by the presence of nearby unstable modes, which are essential for the locking process. Prograde modes are typically excited at high orbital frequencies, where the tidal amplitudes remain large, even far from resonance, and exceed the threshold required for resonance locking. Importantly, unstable prograde modes themselves cannot produce locking. Their influence is transient and instead helps drive the system into a stable locking configuration. While prograde locking also results in angular momentum deposition, the rate is typically two orders of magnitude lower than that of retrograde modes. As a result, stellar spin evolution in this regime is primarily governed by structural expansion. Nevertheless, depending on the rotation rate, age, and internal structure, locking can also occur in low-amplitude unstable prograde modes, leading instead to a slow angular momentum loss via tides. 

\subsection{Numerical resolution of the problem}\label{subsect_numerical_resolution}
\begin{figure*}[]
\centering
\includegraphics[width=15cm]{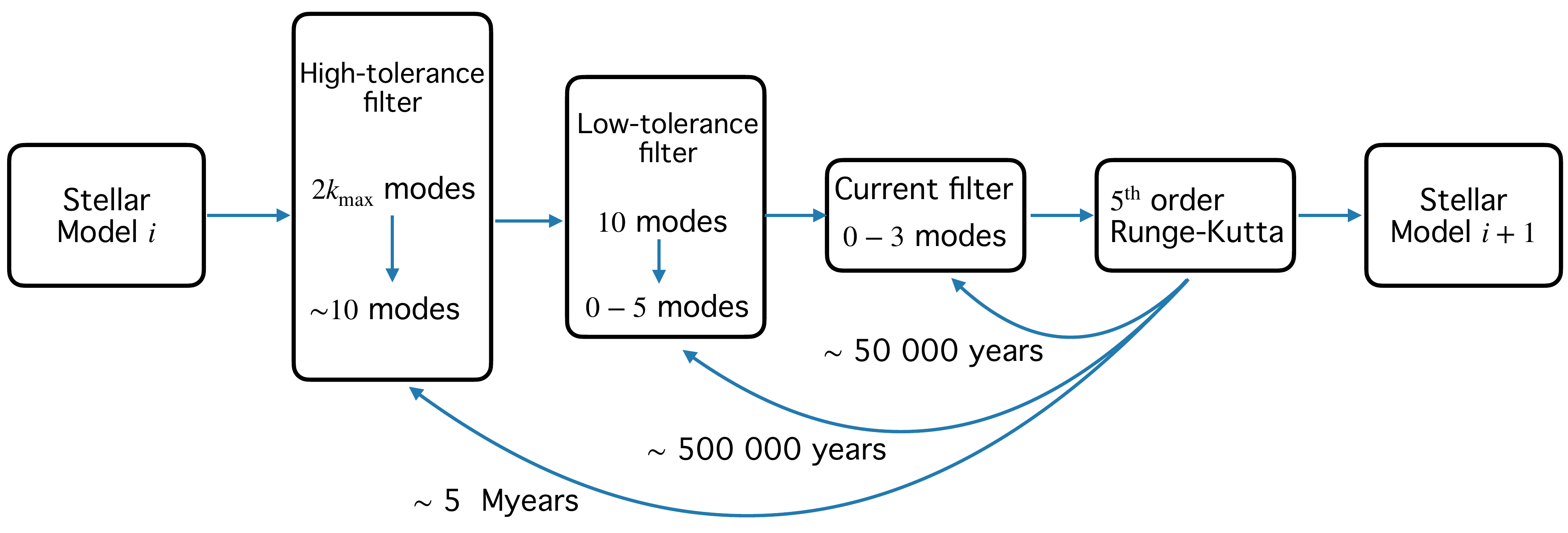}
\caption{Illustration of the principle behind the multi-layer filter system used to compute the dynamical tides with the on-the-fly solving method for a $\gamma$-Doradus star with a mass of $1.5M_\odot$.}
\label{fig.operating_diagram}
\end{figure*}

While the physical principle behind orbital evolution codes properly including the effect of dynamical tides has been known for a long time (see \cite{Smeyers1998,Willems2002,Witte1999a,Witte1999b,Witte2001,Savonije2002,Willems2002,Willems2003,Savonije2005,Willems2010}), the real challenge in developing such a tool is numerical, particularly in terms of computation time. Following the orbital evolution of binary systems induced by a single resonance or a few resonances was, for example, already accomplished in \cite{Willems2002,Witte1999a,Witte1999b,Witte2001,Savonije2002,Savonije2005,Fuller2021}. However, tracking and including all the modes impacting the orbital evolution of the system at a given time is particularly challenging.
\\~\\Four quantities are integrated with our method: $a$, $e$, and the angular momentum of the two stars. We thought that the best way to integrate these quantities was through what we refer to as the `scan method'. As introduced in \autoref{subsect_formulation_dy_tides} and Eqs. \ref{eq_tidal_perturbation} and \ref{eq_norm_Im_Flmk}, a normalised perturbation from the companion can be introduced to obtain the normalised tidal response ($\mathrm{Im}(F_{\ell,m,k}) $), independent of the orbital parameters of the system. Then, scans of the normalised tidal response can be performed and stored as a function of the forcing frequency. The properties of a mode at a given time and given forcing frequency are then obtained by interpolating between scans obtained at a given time step (dictated by the stellar model) and the following time step of the stellar model. The main advantage of this methodology is that all the heavy computations are performed outside the orbital evolution code, and once the scans are ready, the integration over the orbit can be quite fast.
\\~\\We developed a version of the orbital evolution code exploiting this methodology; however, we faced various challenges that greatly limited its precision and computing time. The assumption that the problem is linear is not entirely justified when the rotation of the stars is included. It is well known that the effect of rotation on the frequencies of free oscillations is non-linear \citep{Unno1989}. Therefore, assuming a linear effect of rotation or totally neglecting it is not possible, except for slowly rotating stars. This led us to add an extra dimension in the scan of the oscillation spectrum. If we want to include the effect of rotation, we need to scan the oscillations both through the evolution of age and rotation. The main problem then became computation time. By adding a dimension to the scans, we drastically increased the computation time, as we needed to cover the entire parameter space in rotation at each time step of the model, effectively multiplying the scan time by the number of rotation steps chosen. Another concern was the interpolations themselves. While it was not a problem to develop interpolation methods that were reliable in the parameter space (age, rotation), many questions remained regarding the validity of interpolating oscillation modes, particularly in cases where modes or regions exhibit changes in behaviour (damping or driving) along with age or rotation.
\\~\\To mitigate these drawbacks, we developed a method exploiting the advantages of the direct modelling method \citep{CAFeiN2013,gyre_tides2023,Fellay2025}, which we refer to as `on-the-fly computation'.  The principle of this method is to compute the dynamical tides on the fly using our stellar oscillation code for a very limited number of modes. Instead of pre-computing a large number of scans for modes that are not necessarily used, we focused only on the dominant modes at a given time. The main challenge of this methodology is developing a highly efficient algorithm capable of detecting which oscillation modes to ignore and where the integration time step can be increased. We discuss the performance issues in more detail in \autoref{subsect_performances}. In this section, we focus on the global algorithm we used to integrate the tidal evolution equations on the fly.
\\~\\Optimising the integration algorithm is central to reducing computation time. The primary objective of this optimisation is to control and monitor the integrator truncation errors while avoiding unnecessary oversampling of the dynamical tides. To achieve this, we used a fifth-order Runge-Kutta algorithm with step doubling coupled with a fourth-order Runge-Kutta algorithm to monitor truncation errors \citep{Numerical1986}. This setup enables an adaptive time step designed to keep truncation errors below a specified threshold by adjusting the integration step size accordingly. This algorithm is a standard approach for integrating the effects of tides on planetary or binary system orbital evolution \citep{Jiang2010,Bolmont2015,Luna2020}. It ensures efficient integration with minimal sampling of the forced oscillations while maintaining high precision, even near resonances. Additionally, we imposed an artificial upper limit on the time step to ensure that no sharp resonance contributions were missed. We also explored other integration methods, such as the Bulirsch-Stoer integration method \citep{Numerical1986}, that are standard integrators to compute the evolution of planetary systems \citep{Chambers1999,Mardling2002,Fabrycky2014}. However, these methods are rarely well-suited for functions with sharp transitions, such as tidally excited oscillations \citep{Numerical1986}.
\\~\\As previously mentioned, the on-the-fly method requires drastically reducing the number of oscillation modes tracked. To ensure that the minimum number of modes are integrated while maintaining high precision, we developed a multi-layer filter, the principle of which is described in \autoref{fig.operating_diagram}.
This multi-layer filter contains three filters, each called at different frequencies. The high tolerance filters, which are computationally expensive, are called less often than the low tolerance filters, which are less computationally demanding. When integrating the orbital evolution of the system over the time span between two successive stellar models, the first filter applied to decide which modes are worth tracking is the high tolerance filter. In this step, the impact of dynamical tides at a given time is computed using our stellar oscillation code for each mode with a sufficiently low radial order (for all the computation in this article, we considered modes with a radial order up to $n=200$).  The exact upper limit in terms of co-rotating frequency depends on the mode type.  In general, around $2k_{\mathrm{max}}$ (where $k_{\mathrm{max}}$ is the maximum number of orbital harmonics to consider, which varies as a function of the eccentricity due the shift of the peak in amplitude of the Hansen coefficients depending on $k$ \citep{Piotr2023}) modes were computed at this step. We then selected only the modes for which the orbital and rotational characteristic timescales (as given in \autoref{eq_characteristic_timescales}) are below a given threshold, which is quite high (hence the high tolerance), and arbitrarily taken as a thousand times the nuclear timescale of the system. Only the few selected modes are passed to the low-tolerance filter, where we computed, in addition to the characteristic timescales, their derivatives with respect to time to detect which oscillation modes are close to resonance (which can drastically increase their effect on the orbital and rotational parameters). Only modes with sufficiently low characteristic timescales (which were arbitrarily chosen to be a hundred times the stellar nuclear timescale) or high characteristic timescale derivatives were selected and saved.  For the few modes selected through this process, a final, frequently updated filter was applied based only on the current characteristic timescales (arbitrarily chosen to be ten times the stellar nuclear timescale),  to ensure that only the dominant modes are integrated at a given time step. This multi-layer system is directly embedded in the Runge-Kutta algorithm to avoid wasting computation time. The refresh frequencies of all filters were parameterised based on the nuclear timescale. 

\subsection{Performances and precision}\label{subsect_performances}
As discussed previously, the principal difficulty in fully including dynamical tides into orbital evolution codes is the computational time. In this section, we examine the performance and precision of the two solving methods presented in \autoref{subsect_numerical_resolution}. For both methodologies, the execution time is mainly dictated by the number of calls to MAD_tides, which represents by far the most computationally expensive area of the code. In the scanning method, the number of calls to the oscillation code cannot be significantly reduced without decreasing precision. The main drawback of this methodology is that during the scans, a large portion of the computations remains unused, but this cannot be avoided.
\\~\\For the scans, we determined that around 200 sampling points (and uses of the oscillation code) per resonance are sufficient to guarantee an acceptable resolution of the function to allow for precise interpolations. When accounting for rotational effects, we needed to sample four types of modes (one for each $m$, plus Rossby modes). If we focus solely on low-radial-order modes (expected to have the highest amplitudes and to be unstable), which we assumed to be those with radial orders up to $n=100$, we end up with 200 modes to sample at each stellar age and rotation rate. Over the course of an entire stellar evolution sequence, requiring 300 stellar models (for a $\gamma-$Doradus or SPB star), we must call our stellar oscillation code approximately 24 million times (solving 24 million instances of the forced non-adiabatic oscillation differential equation system). After optimising and parallelising MAD_tides (using a processor with 24 cores (eight performance cores plus 16 efficiency cores) and 32 threads), we achieved a computing speed of approximately 12000 forced oscillations per minute. This means that scanning the oscillations of an evolutionary sequence takes about 34 hours, plus an additional five hours are required to identify the free resonance frequencies. Ultimately, each scan for a given rotation requires approximately an equivalent of 27.5 million calls to the forced stellar oscillation code. To encompass the possible initial stellar rotation rates, at least two such scans with different rotation rates are necessary, bringing the total to around 55 million calls to our stellar oscillation code. Additionally, integrating the system’s evolution requires several hours due to the extensive number of interpolations, although we note that this part of the code has not been fully optimised. The objective of the on-the-fly method is to drastically reduce the computational cost compared to the scan method or to increase the precision for the same cost.
\\~\\ With on-the-fly computation, many quantities can be precomputed and stored, leaving only the problem of solving the forced oscillation differential equation system. The primary advantage of this approach is that it significantly reduces the number of required forced oscillation computations. This efficiency is achieved by the algorithm itself and, in particular, the various filters presented in \autoref{subsect_numerical_resolution} and their combinations. Since the number of modes included in the computation is determined dynamically during integration rather than being fixed, an exact estimation of computational time is not feasible and depends on the number of modes being tracked. However, we estimate that the computation time is roughly eight hours per billion years if only a few modes require integration. 
\\~\\For $\gamma$-Doradus stars, which have a typical nuclear timescale of approximately three billion years, the on-the-fly method costs already less than a single scan from scanning method per integration and achieves a higher overall precision. For SPB stars, which have nuclear timescales on the order of a few hundred million years, multiple integrations using the on-the-fly method cost less than a single scan. It should be noted that the scan method can operate at a lower overall precision than the on-the-fly method.  The advantage of the scan method is that once the scan are done,  multiple low cost integrations can be performed with various initial orbital parameters.  However, most errors and imprecisions arise from interpolation and scan resolution (rather than the integrator itself), making them difficult to quantify. It is also worth noting that for rapid rotators, a large number of scans may be required to precisely interpolate the effects of rotation.

\section{Full integration of the effect of dynamical tides}\label{sect_intergration_dy}

Before presenting our results, it is important to highlight that the theory of the dynamical tides developed by \cite{Zahn1975,Zahn1977} is not expected to remain valid in the forcing frequency or stellar rotational regimes considered in this article.  In particular, in the regime of low-radial-order modes the condition $\sigma\tau_{\mathrm{dyn}}\ll 1$ is not fulfilled and the proportionality of the tidal dissipation with respect to the forcing frequency studied by \cite{Zahn1975,Zahn1977} does not apply.  Moreover,  in fast rotating stars, the rotation has a strong effect on the oscillations eigenfunctions,  which was not considered by \cite{Zahn1975,Zahn1977}.  In the limit of low forcing frequencies ($\sigma\tau_{\mathrm{dyn}}<0.1$), we verified that TREMOR’s results are consistent with the prescriptions of \cite{Zahn1975,Zahn1977}.  We found that TREMOR  and Zahn’s prescription yields the same order of magnitude for dynamical tidal effects in this low forcing frequencies  regime.  However, a more detailed study is needed to fully explore the synergies and limitations of our approach compared to Zahn’s prescription. In this article, we focus exclusively on the regime where \cite{Zahn1975,Zahn1977} is not expected to be valid or dominate, specifically, cases with high forcing frequencies ($\sigma \tau_{\mathrm{dyn}}>0.1$), where low-radial-order oscillation modes are excited by tidal interactions.

\subsection{Reference model for this study}\label{subsect_reference_model}
Our goal is to investigate the impact of both stable and unstable oscillation modes on the orbital and rotational evolution of binary systems. For this reason, we focus on stars within the $\delta$-Scuti, $\gamma$-Doradus, and SPB instability strips, as their evolution, modelling, rotation, and pulsation properties are well studied \citep{Townsend2005,Bouabid2013,Li2019,Li2020,Saio2021}.  Due to the assumptions made in our stellar oscillation code, particularly the use of the traditional approximation, we were unable to reliably compute oscillation properties for stars with deep convective envelopes \citep{Ogilvie2004,Ogilvie2007} or for pressure modes. As a result, our study is limited to stars within the $\gamma$-Doradus and SPB instability strips. For this study, we chose one reference sequence of evolution of $\gamma-$Doradus  stars. This sequence of evolution was computed with the Code Liégeois d'Évolution Stellaire \citep[CLÉS;][]{Scuflaire2008a}. We chose a sequence in the lower end of the mass regime, and we chose a star with a mass of $1.5\ M_\odot$.  For this model we used an initial hydrogen mass fraction of $X=0.72$, a solar metallicity $Z=0.015$, a mixing length convective parameter of $\alpha=2.0$, and a step-like overshooting of $\alpha_{\mathrm{over}}=0.2$. The model was stopped when the central hydrogen mass fraction reached $X_C=0.05$, and we reached a final age of $2.7$ Gyr for this evolutionary sequence. The parameters of the reference model were chosen to have unstable modes at the zero age main sequence and during a significant part of the main sequence. For all the stellar modelling, we used the AGSS09 abundances \citep{Asplund2009}, the FreeEOS equation of state \citep{Irwin2012}, the OPAL opacities \citep{Iglesias1996}, the $T(\tau)$ relation from Model-C of \citet{Atmosphere1981} for the atmosphere, the mixing length theory of convection implemented as in \citet[][]{Cox1968}, and the nuclear reaction rates of \cite{Reaction2011}.  
\\~\\To highlight the impact of resonance locking (or tidal forces in general) on the evolution of stellar rotational properties in our models, we compared them to the evolutionary tracks of isolated stars (i.e. assuming no tidal interactions at all) with the same initial parameters.  Since CLÉS does not include a treatment of angular momentum transport, we computed reference models with Modules for Experiments in Stellar Astrophysics \citep[MESA, version 24.08.1][]{Paxton2011, Paxton2013, Paxton2015, Paxton2018, Paxton2019, Jermyn2023}, compiled using the MESA Software Development Kit for Linux \citep[version 24.7.1][]{richard_townsend_2024_13768913}, which accounts for those processes. We then compared the surface velocities predicted by MESA models (considering only single stars) with those obtained from our CLÉS models including the effects of dynamical tides. This comparison enabled us to evaluate whether tidal effects can produce observationally significant differences. To ensure maximum consistency with the simulations run using TREMOR, we adopted the same initial stellar mass, chemical composition, opacity tables,  mixing length and overshooting prescriptions, and termination condition. After evolving a non-rotating pre-main sequence model up to just before the zero age main sequence, we relaxed the model into solid-body rotation with a rate matching that in the TREMOR simulations. We then allowed differential rotation to develop between stellar layers, assuming shellular approximation. All five rotational mixing processes implemented in MESA were included, along with the associated angular momentum diffusion in the radial direction. These rotationally induced mixing mechanisms include dynamical shear instability, Solberg-H\o{}iland instability, secular shear instability, Eddington-Sweet circulation, and the Goldreich-Schubert-Fricke instability. A detailed description of how rotation and associated mixing are treated in MESA can be found in \cite{Paxton2013} and \cite{2000ApJ...528..368H}. In our MESA models, we neglected stellar winds, which are negligible for the stellar mass regime of interest here. Atmospheric boundary conditions were taken from \cite{2003IAUS..210P.A20C} model atmospheres, and convective boundaries were determined using the Ledoux criterion with the so-called convective premixing algorithm \citep[][their Sect. 5.2]{Paxton2019}. The micro- and macrophysics data used by the MESA modules are briefly summarised in the Appendix (Sect. \ref{app:mesa}).

\subsection{Fast-rotating stars (retrograde modes)}
In $\gamma$-Doradus stars, a large instability strip exists due to the convective blocking, where the flux is periodically blocked at the base of the near surface convective zone \citep{Guzik2000, Dupret2005}. This extended instability strip results in a significant fraction of overstable oscillation modes, which can create resonance locking. We divided this study of $\gamma$-Doradus stars into two parts. First the cases of fast-rotating stars dominated by the effect of retrograde modes is described in this section, and then the case of slow-rotating stars in tight orbits, where the tidal interactions are dominated by prograde modes, is described in \autoref{subsect_prograde_modes_TREMOR}.
\subsubsection{Low-eccentricity orbital evolution}
\begin{figure}[h]
\centering
\includegraphics[width=\hsize]{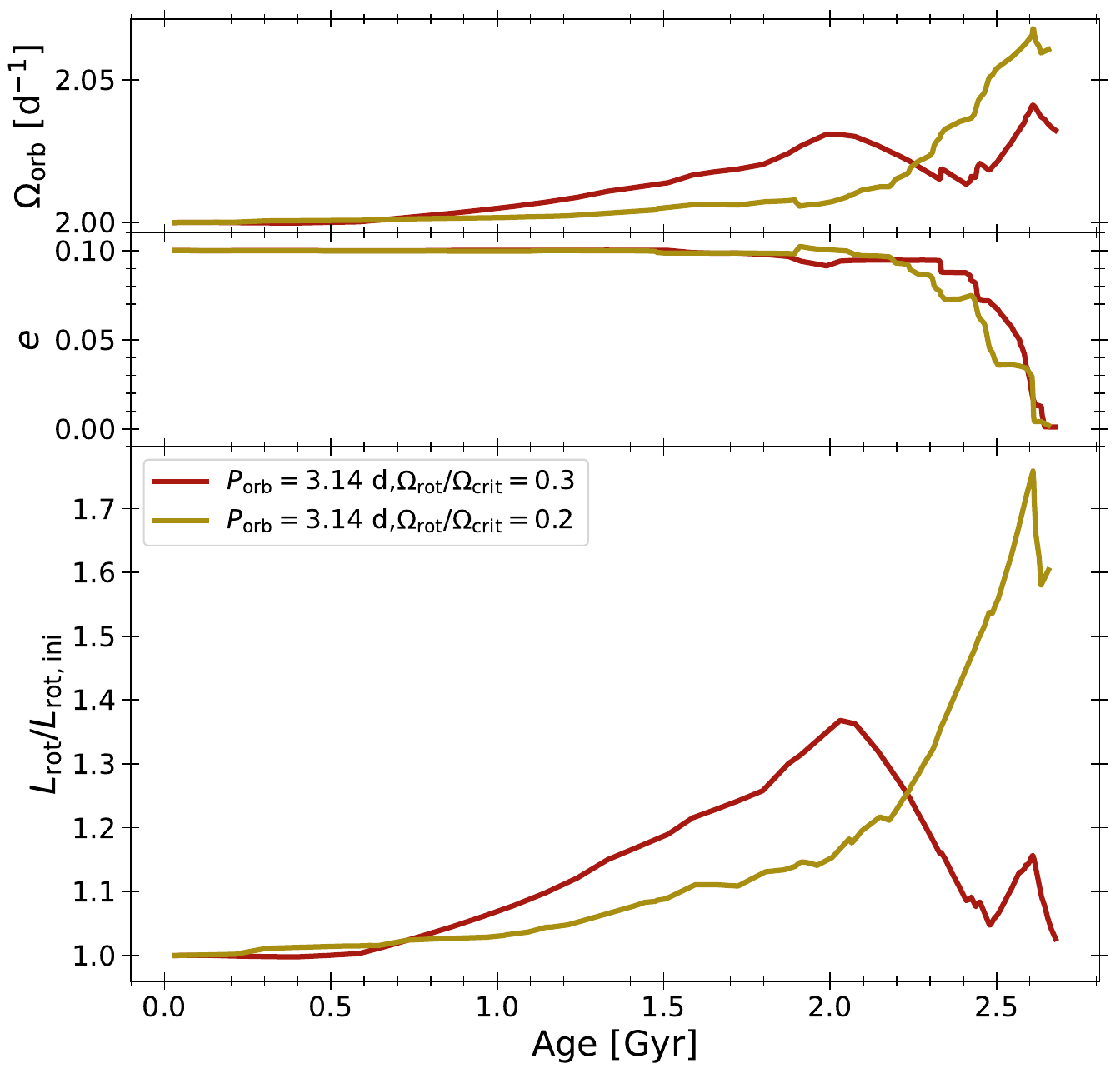}
\caption{Evolution of the orbital frequency (top panel), eccentricity (middle panel), and stellar angular momentum (bottom panel) as a function of system age obtained with TREMOR. Two initial stellar rotation rates were considered: $\Omega_{\mathrm{rot}}/\Omega_{\mathrm{crit}} = 0.2$ and $0.3$. The initial eccentricity was set to $0.1$, and the initial orbital period of the system is $3.14$ days.}
\label{fig.evolution_e0.1_hrot}
\end{figure}

\begin{figure}[h]
\centering
\includegraphics[width=\hsize]{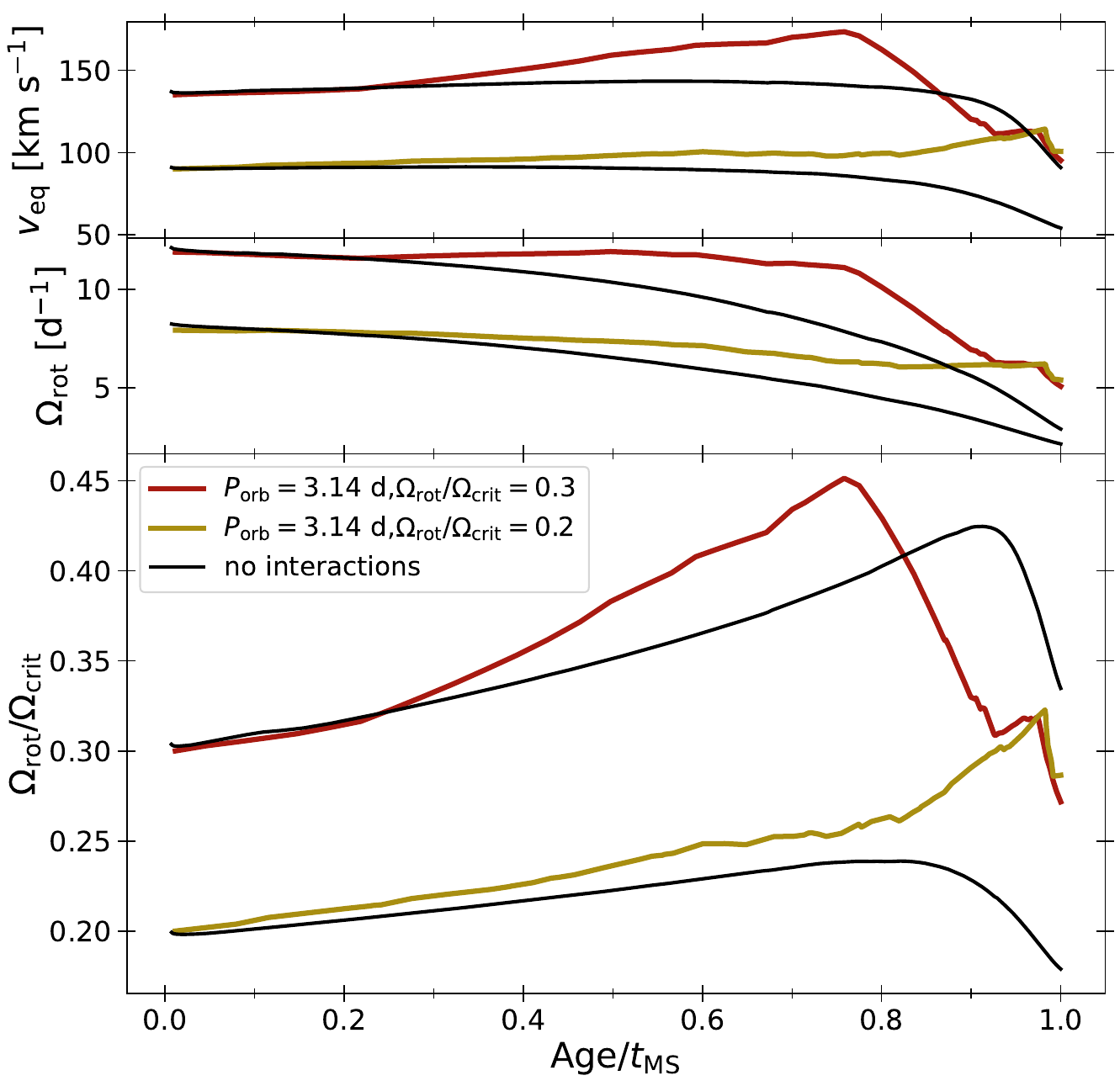}
\caption{Evolution of the surface rotation velocity (top panel), the stellar rotation rate normalised by the critical rotation rate (middle panel), and the stellar angular momentum (bottom panel) as a function of system age, as computed with TREMOR. Two initial stellar rotation rates are considered: $\Omega_{\mathrm{rot}}/\Omega_{\mathrm{crit}} = 0.2$ and $0.3$. The initial eccentricity was set to 0.1, and the initial orbital period is 3.14 days. The black track corresponds to the evolution predicted by single-star models.}
\label{fig.evolution_e0.1_hrot_adrot}
\end{figure}

For low-eccentricity systems ($e < 0.3$), the problem of dynamical tides is significantly simplified. In such cases, only the first few orbital harmonics ($k$) can effectively excite oscillation modes with an amplitude sufficient to impact the orbital and rotational evolution of the star. This reduces the number of modes that must be tracked and minimises mode-mode interactions. We explored the effect of dynamical tides on the orbital and rotational evolution of the system by considering a twin system composed of two $1.5 M_\odot$ $\gamma$-Doradus stars with an initial orbital angular frequency of $\Omega_{\mathrm{orb}} = 2$ rad d$^{-1}$, initial stellar rotation rates of $\Omega_{\mathrm{rot}}/\Omega_{\mathrm{crit}} = 0.2, 0.3$, and an eccentricity of $0.1$. In \autoref{fig.evolution_e0.1_hrot}, we illustrate the resulting evolution of the stellar angular momentum, semi-major axis, and eccentricity. Additionally, \autoref{fig.evolution_e0.1_hrot_adrot} presents the evolution of the rotation rate, rotation to critical rotation ratio, and surface rotation velocity for the same systems.
\autoref{fig.evolution_e0.1_hrot} shows that throughout most of the evolution, angular momentum is deposited into the stars,  as indicated by $L_{\mathrm{rot}}/L_{\mathrm{rot,ini}}>1$ in panel 3 of \autoref{fig.evolution_e0.1_hrot}. This deposition occurs due to the inverse tides.
For the system with the fastest initial rotation, the deposition of angular momentum reaches a maximum increase of 35$\%$ relative to its initial spin angular momentum at 2 Gyr, before undergoing a rapid decline due to the influence of stable oscillation modes, which have high amplitudes given the low orbital period. This sudden drop is an evolutionary effect, as most of the unstable modes present in the early main sequence transition into stable modes in later stages. For an initial rotation of $\Omega_{\mathrm{rot}}/\Omega_{\mathrm{crit}} = 0.2$ the increase  of angular momentum reaches $70\%$ of the initial angular momentum at the end of the main sequence.  As expected, the effect of this angular momentum exchange on the orbital parameters is minimal since the orbital moment of inertia is significantly larger than the stellar moment of inertia, except near the end of the main sequence, where stable modes efficiently damp the eccentricity. To further investigate the rapid angular momentum deposition, \autoref{fig.evolution_e0.1_hrot_modes_details} illustrates the evolutionary timescales associated with angular momentum exchange, broken down by individual oscillation modes. The figure shows that for most of the system’s evolution, a single dominant retrograde mode ($\ell=2, m=2, k=-2, n=27$ for $\Omega_{\mathrm{rot}}/\Omega_{\mathrm{crit}} = 0.2$ and  $\ell=2, m=2, k=-2, n=17$ for $\Omega_{\mathrm{rot}}/\Omega_{\mathrm{crit}} = 0.3$) is responsible for long-term resonance locking and sustained angular momentum deposition. Due to the stability of resonance locking and the small number of high-amplitude stable oscillation modes (resulting from the system’s low eccentricity), mode crossings are rare throughout the evolution.

\subsubsection{High-eccentricity orbital evolution}\label{subsubsect_high_eccentricty evo}
\begin{figure}[h]
\centering
\includegraphics[width=\hsize]{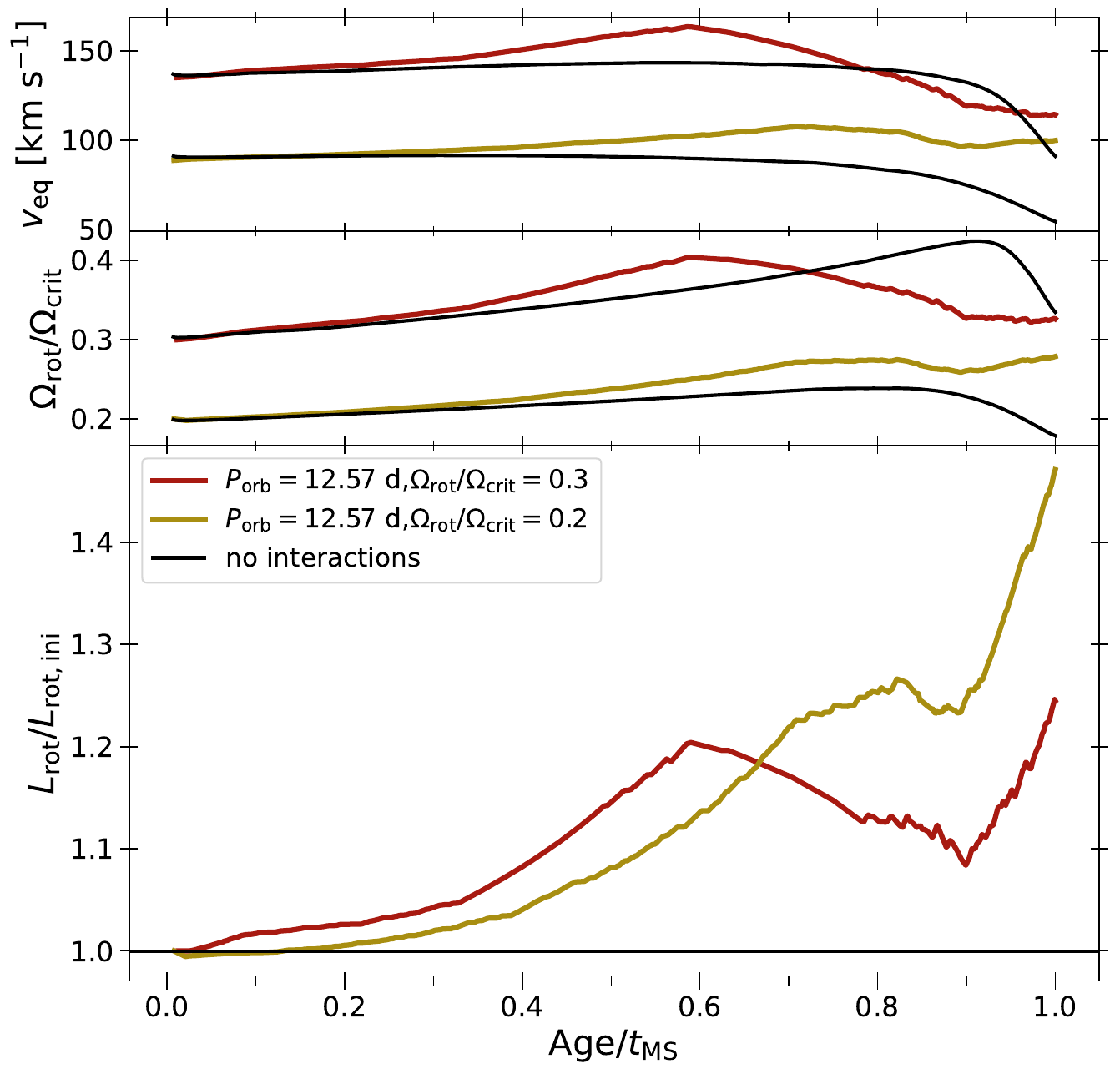}
\caption{Evolution of the surface rotation velocity (top panel), the stellar rotation rate over the critatical rotation rate   (middle panel), and stellar angular momentum (bottom panel) as a function of system age obtained with TREMOR. Two initial stellar rotation rates were considered: $\Omega_{\mathrm{rot}}/\Omega_{\mathrm{crit}} = 0.2$ and $0.3$. The initial eccentricity was set to $0.55$, and the initial orbital period of the system is $12.57$ days. The black track corresponds to the evolution predicted by single stars models.}
\label{fig.evolution_e0.55_hrot}
\end{figure}

In the case of high-eccentricity binaries, more resonances with oscillation modes are expected, and resonance locking is likely to be shorter-lived.  We explored the effect of dynamical tides on the orbital and rotational evolution of the system by considering a twin system composed of two $1.5 M_\odot$ $\gamma$-Doradus stars with an initial orbital angular frequency of $\Omega_{\mathrm{orb}} = 0.5$ d$^{-1}$, initial stellar rotation rates of $\Omega_{\mathrm{rot}}/\Omega_{\mathrm{crit}} = 0.2, 0.3$, and an eccentricity of $0.55$. In \autoref{fig.evolution_e0.55_hrot}, we illustrate the evolution of angular momentum exchange induced by dynamical tides and its impact on stellar rotation.
The same behaviour as in the low-eccentricity case is exhibited in \autoref{fig.evolution_e0.55_hrot} but amplified. In particular, for the case with an initial rotation of $\Omega_{\mathrm{rot}}/\Omega_{\mathrm{crit}} = 0.2$, the stellar angular momentum increases by 50$\%$. In \autoref{fig.evolution_e0.55_hrot_modes_details}, we further illustrate the timescales of angular momentum exchange for each tracked oscillation mode. When comparing Figs.~\ref{fig.evolution_e0.1_hrot_modes_details} and \ref{fig.evolution_e0.55_hrot_modes_details},  we observed that while for low-eccentricty cases, a single resonance locking is responsible for all the deposition of angular momentum, for higher-eccentricty cases a succesion of short-lived locking occurs. Despite their short-term nature, these lockings effectively sustain resonance locking throughout most of the main sequence. The primary difference in the efficiency of angular momentum deposition between low- and high-eccentricity systems is determined by the probability of achieving a resonance lock. In high-eccentricity systems, the likelihood of encountering a resonance capable of locking the system is significantly higher. This is because unstable oscillation modes can be excited by a greater number of orbital harmonics, allowing the system to rapidly re-lock after unlocking.
\subsection{Close-in slow-rotating  systems (prograde modes resonances)}\label{subsect_prograde_modes_TREMOR}
\begin{figure}[h]
\centering
\includegraphics[width=\hsize]{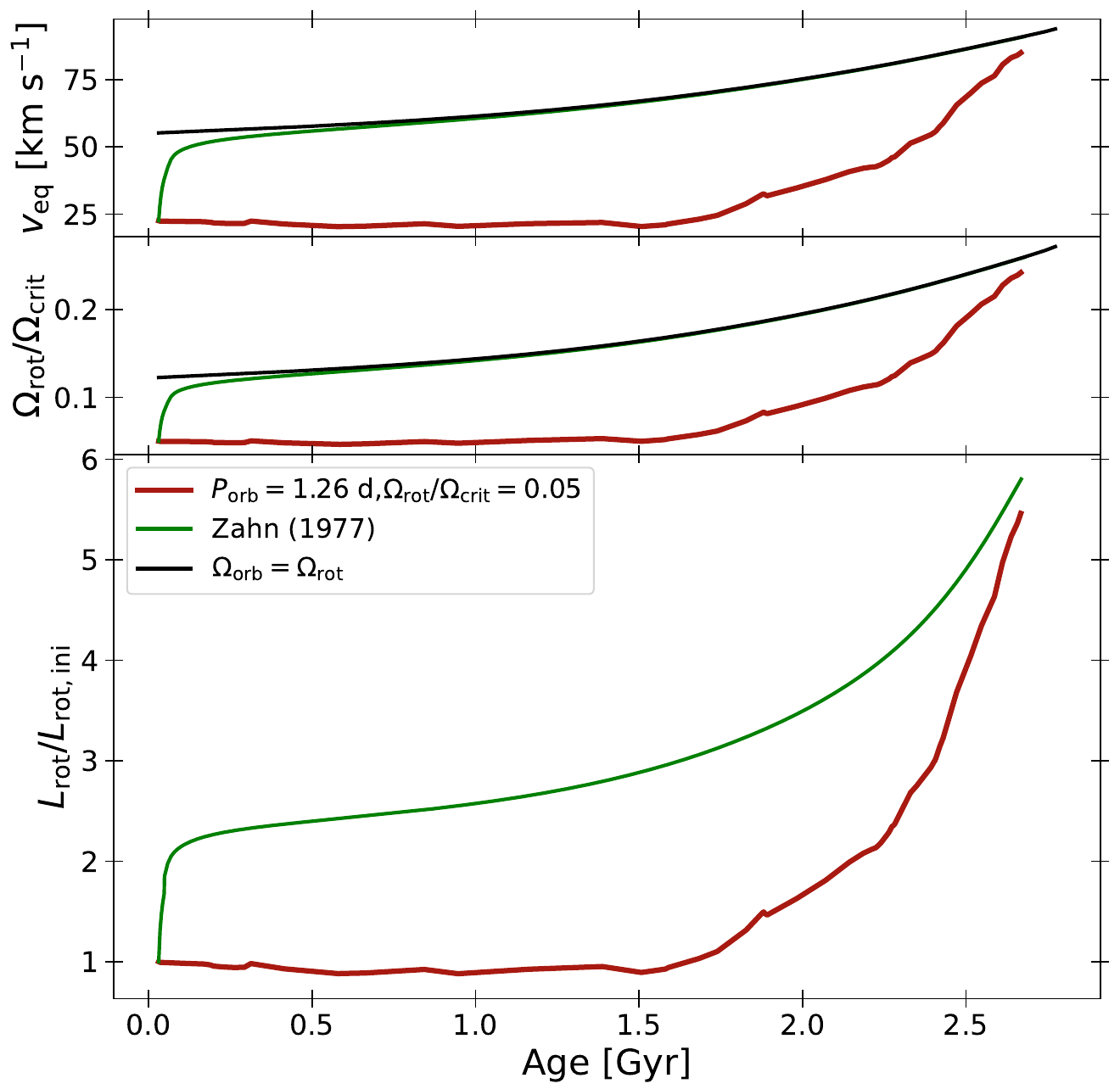}
\caption{Evolution of the surface rotation velocity (top panel), the stellar rotation rate over the critical rotation rate   (middle panel), and stellar angular momentum (bottom panel) as a function of system age obtained with TREMOR. The initial stellar rotation rate is $\Omega_{\mathrm{rot}}/\Omega_{\mathrm{crit}} = 0.05$, the initial eccentricity was set to $0.1$, and the initial orbital period of the system is $1.26$ days. The green track corresponds to the evolution predicted with the formalism from \cite{Zahn1975,Zahn1977}. The black line corresponds to the rotation frequency for which the system is pseudosynchronised. }
\label{fig.evolution_e0.1_lrot}
\end{figure}

For prograde modes, we required that positive forcing frequency for all $k>0$   ($2\Omega_{\mathrm{rot}} < \Omega_{\mathrm{orb}}$) to ensure a positive forcing frequency. Furthermore, to study unstable prograde modes, we needed to reach high forcing frequencies (i.e. $\sigma \tau_{\mathrm{dyn}} > 0.1$). Consequently, we focused on close-in systems with slowly rotating components. For the same twin system composed of two $1.5 M_\odot$ $\gamma$-Doradus stars, we considered the case where the initial orbital angular frequency is $\Omega_{\mathrm{orb}} = 5$ d$^{-1}$, the initial stellar rotation rate is $\Omega_{\mathrm{rot}}/\Omega_{\mathrm{crit}} = 0.05$, and the eccentricity is $0.1$. \autoref{fig.evolution_e0.1_lrot} illustrates the evolution of angular momentum exchange induced by dynamical tides and its impact on stellar rotation for this system.
\autoref{fig.evolution_e0.1_lrot} also shows that even at low orbital frequency, the stellar rotation is not significantly affected by dynamical tides during the first $\sim 1.5$ Gyr of the evolution.  Subsequently, the dynamical tides have their classical dissipative effect, and the system is driven towards synchronisation.
In the same figure, we include a track based on the prescription from \cite{Zahn1975,Zahn1977}, which suggests that given the separation, the system should synchronise within approximately 500 Myr. Examining the amplitude of each oscillation mode forced by tides (illustrated in \autoref{fig.evolution_e0.10_lrot_modes_details}), we observed that multiple modes produce simultaneous short-timescale resonances. However, their net effect on rotation remains negligible. In this particular case, we identified an instance of four modes ($\ell=2,m=-2, k=1,2,3,4$ with various radial orders changing through time) resonance locking, which decreases the angular momentum of the star. For this locking to occur, we required the presence of multiple modes with opposite signs that become locked together for extended periods. Across our simulations, we consistently observed that resonance locking is particularly stable. When a resonance locking event occurs with a mode, the relative motion of oscillation mode frequencies slows down significantly, or in some cases, their relative motion becomes fixed because the rotation frequency is nearly frozen. This form of resonance locking exploits such a behaviour to maintain stability over long timescales. Given our results, a mathematical explanation of this phenomenon is necessary to determine why certain resonances are particularly stable and up to which orbital frequency this phenomenon can occur.

\subsection{Stability of resonance locking}
We initially assumed that different oscillation modes would evolve at different rates, making resonance locking stable only over short timescales, limited by the time it takes for a stronger or faster-moving resonance to either unlock or capture the system. However, if we express the condition for resonance locking as
\begin{equation}
\dfrac{\mathrm{d}\sigma}{\mathrm{d}t}=k\dfrac{ \mathrm{d}\Omega_{\mathrm{orb}}}{\mathrm{d} t}+m\dfrac{\mathrm{d}\Omega_{\mathrm{rot}}}{\mathrm{d} t},
\end{equation}
it can be observed that if $\dfrac{\mathrm{d}\sigma}{\mathrm{d}t}$ does not vary significantly for modes with different radial orders and given that $k\dfrac{ \mathrm{d}\Omega_{\mathrm{orb}}}{\mathrm{d} t}\ll m\dfrac{\mathrm{d}\Omega_{\mathrm{rot}}}{\mathrm{d} t}$, then when this condition is satisfied for one mode, it is also satisfied for modes with nearby radial orders. In such a scenario, the relative positions of resonances become effectively fixed over time or evolve much more slowly than under normal conditions. For `high' radial order modes ($n>20$), which is the typical range where unstable modes are found in $\gamma$-Doradus stars \citep{Bouabid2013}, this condition is met. As a result, resonance locking can be sustained over timescales comparable to the stellar main sequence since resonance crossings (which could disrupt the locking) become much less frequent. This behaviour was first identified in our numerical simulations using TREMOR (see  \autoref{sect_intergration_dy}), confirming that it is a robust phenomenon.

\section{Discussion and conclusion}\label{sect_discussion}
In this study, we have developed, for the first time, a numerical method called TREMOR (introduced in \autoref{sect_TREMOR}), capable of integrating the simultaneous impact of all stellar oscillation modes excited by the presence of a companion on the orbital and rotational evolution of a system. Using TREMOR, we performed numerical integrations for twin binary systems with various initial parameters. Our results show that if the rotation is above approximately $15\% $ of the critical rotation, resonances with unstable retrograde modes in the co-rotating frame and prograde in the observer frame lead to to an increase in stellar rotation and the deposition of angular momentum into the star. We find that resonance locking is a particularly stable state that can persist over long timescales. Throughout the main sequence, resonance locking with inverse tides can increase stellar angular momentum by up to 70$\%$ in both low- and high-eccentricity systems.
\\In close-in systems ($\Omega_{\mathrm{orb}} \tau_{\mathrm{dyn}}\gtrsim 0.1$) with slowly rotating components ($\Omega_{\mathrm{rot}} \tau_{\mathrm{dyn}}\lesssim 0.05$), prograde modes can also induce resonance locking. The effect of this locking is to enforce a slow evolution of stellar rotation, preventing the rapid synchronisation expected in close-in systems. In regimes where the forcing frequencies are small compared to the inverse of the dynamical timescale, we expect tidal dissipation mechanisms described by \cite{Zahn1975,Zahn1977} to dominate.
\\~\\A key distinction between our approach and that of \cite{Fuller2021} is that we account for the evolution of the forcing frequency due to stellar rotation changes, include the effects of rotation on oscillation modes, and incorporate stellar evolution. Additionally, we do not impose a resonance locking. As a result, our findings differ significantly from those of \cite{Fuller2021}. Namely, our results suggest that resonance locking is not a violent process capable of forming Be stars. Instead, it is a low-acting, long-lasting mechanism that can modify the stellar rotation over most of the main sequence. This mechanism allows stars to retain a trace of their initial rotation, even in close-in systems, where strong tidal interactions would otherwise be expected to erase it.
\\~\\One of the primary limitations of our model is the assumption of solid-body rotation. Under this assumption, we demonstrated that resonance locking provides an efficient angular momentum exchange mechanism capable of modifying global rotation. In reality, for high-radial-order oscillation modes (such as those considered in \cite{Zahn1975,Zahn1977}), most of the tidal dissipation likely occurs in the near-surface layers \citep{Goldreich1989}. This leads to surface synchronisation before angular momentum can be deposited into deeper layers.  On the other hand, a few observations \citep{Kallinger2017,Sowicka2017} suggest that large differential rotation can exist in stars that appear synchronised at the surface. This raises the possibility that inverse tides may induce asynchronisation in the deeper stellar interior, while the outer layers remain synchronised due to equilibrium tides or high-radial-order forced oscillation modes. The exact processes governing inverse tidal dissipation remain uncertain. However, in $\gamma$-Doradus stars, the adiabatic approximation breaks down at deeper layers for low-radial-order modes \citep{Dupret2005}, such as our overstable modes. This suggests that dissipation may occur  deeper in the star. Other non-linear mechanisms may also contribute to deep angular momentum deposition via inverse tides. 
Additionally, angular momentum transport processes, such as magnetic torques, are expected to enforce near solid-body rotation \citep{Spruit2002,Fuller2019}, except in the near-surface layers, where a shear can develop. Including differential rotation and its evolution in our models would be a significant improvement, though the physical mechanisms underlying such an implementation remain uncertain, particularly for tidal dissipation. The major results of our study on the evolution of the global rotational angular momentum should not be fundamentally modified by taking into account differential rotation. What remains uncertain is the evolution of the surface rotation.
\\~\\From an observational perspective, Zahn's tidal theory generally agrees with observed surface synchronisation (or pseudo-synchronisation) trends in close-in intermediate-mass star populations \citep{Khaliullin2007}. However, a few observational studies \citep{Zimmerman2017,Lampens2018,Guo2019,Li2020a,Li2020} have reported the existence of slowly rotating close-in binaries.  For instance,  \cite{Zimmerman2017} report that approximately 20$\%$ of their sample consists of strongly sub-synchronous binaries, while the majority are only mildly sub-synchronous. This distribution appears to be inconsistent with the expected tendency towards pseudosynchronous rotation rates predicted by theoretical modelling \citep{Townsend2023}.  However, we did not observe a population of fast-rotating close-in binaries, except for a few outliers or systems where synchronisation may not have been expected to occur yet \citep{Thompson2012,Mantegazza2010,Shporer2016,Li2020a,Saio2022}. For example, \cite{Li2020a} reported six super-synchronous systems located near the transition between synchronised and unsynchronised binaries. However, it remains unclear whether these systems were born as fast rotators and are currently synchronising or whether inverse tides are at play. It should be noted that for fast-rotating stars exhibiting tidally forced oscillations, identifying the oscillation modes and deducing the near-core rotation rate may be challenging.  It is also worth noting that non-linear effects, as investigated by \citet{Fuller2021}, may inhibit or make unlikely resonance locking with retrograde modes since such lockings occur close to resonance, where non-linear effects are expected to be significant. In contrast, lockings involving prograde modes occur away from  resonance, where non-linear effects are expected to be weaker, making them more likely to operate. 
The study of these non-linear effects, however, remains a significant challenge, and current approaches rely on simplified adiabatic modelling.  Further theoretical developments or direct non-adiabatic simulations would be highly valuable for improving our understanding of these phenomena, but this issue is beyond the scope of the present work.
\\~\\In Zahn’s tidal theory, angular momentum deposition is expected to occur primarily near the surface \citep{Goldreich1989}. Thus, classical observation, limited to the outer stellar layers, should be consistent with Zahn’s model provided that low-frequency forcing is present in the system. We also do not exclude other processes, such as equilibrium tides with viscous dissipation \citep{Darwin1880,Zahn1977,Goldreich1977,Zahn1989,Duguid2020,Vidal2020, Barker2021} to still be able to synchronise the surface layers, as $\gamma-$Doradus stars possess an outer thin convective layer. 
If inverse tidal angular momentum deposition occurs deeper within the stellar interior, conventional observational methods (excluding asteroseismology or apsidal motion studies) may fail to detect the rapid rotation of deep radiative zones. Another possible explanation for the lack of observed fast-rotating close-in binaries is that such systems may not form frequently. While fast-rotating single stars are common \citep{Royer2007}, studies such as \cite{Khaliullin2007} and \cite{Li2020a} show that in binary systems, the average stellar rotation rate is significantly lower than that of single stars.
\\Future investigations should focus on detailed asteroseismic studies of pseudo-synchronised binary stars, particularly those with well-characterised surface rotation rates and oscillation frequencies. The key objective of such studies will be to determine whether the near-core rotation is consistent with the observed surface rotation rate. These observations, combined with further theoretical developments, will help refine our understanding of inverse tides, resonance locking, and angular momentum transport in binary stars. 

\begin{acknowledgements} 
L.F was supported by the Fonds de la Recherche Scientifique F.R.S-FNRS as a Research Fellow. This research was supported by the University of Li\`ege under the Special Funds for Research, IPD-STEMA Programme.
 
\end{acknowledgements}
\bibliography{DY_code.bib}
\bibliographystyle{aa}
\appendix
\section{Conservation of the angular momentum in TREMOR}
\begin{figure}[h]
\centering
\includegraphics[width=\hsize]{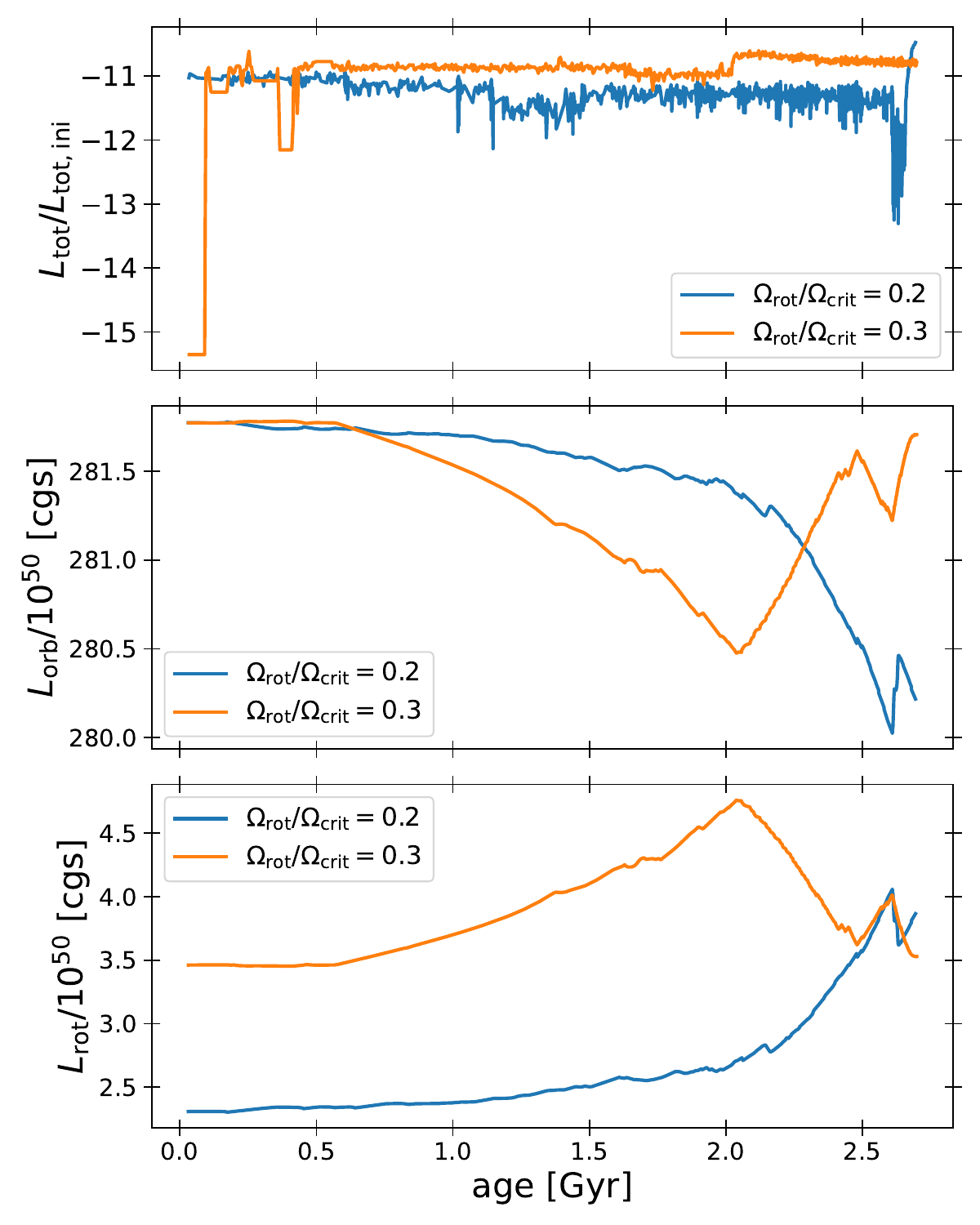}
\caption{The top panel: conservation of angular momentum for the low-eccentricity fast-rotating $\gamma$-Doradus twin system, each line corresponds to a different initial rotation.  The middle panel shows the evolution of the orbital angular momentum in cgs as a function of the time and the lower panel shows the evolution of the evolution of the total rotational angular momentum.}
\label{fig.conservation_angular_momentum}
\end{figure}

In the case where only tidal effects are considered, the total angular momentum of the binary system must be conserved throughout its evolution. Ensuring angular momentum conservation is therefore a crucial test when numerically integrating the effects of tides.
\\~\\With the Runge-Kutta integrator used in TREMOR, a gradual accumulation of numerical errors leading to slight angular momentum non-conservation is expected due to truncation errors. However, we control this accumulation through our adaptive time step, which ensures that the truncation error remains below a predefined threshold. For all computations presented in this article, we set this threshold to $10^{-10}$ for each integrated quantity, namely, the semi-major axis, the eccentricity, and the stellar rotational angular momentum of each star in the system.
\\~\\For the evolution of the low-eccentricity fast-rotating $\gamma$-Doradus twin system, whose evolution is illustrated in \autoref{fig.evolution_e0.1_hrot}, we analyse the conservation of angular momentum in \autoref{fig.conservation_angular_momentum}.
\\~\\As shown in these figures, angular momentum conservation is maintained with a precision of $\sim10^{-10}$. In our case, no cumulative discrepancy arises over time, as the dynamical effects are non-monotonous.

\section{Additional figures for low-eccentricty fast-rotating stars}
\begin{figure}[h]
\centering
\includegraphics[width=\hsize]{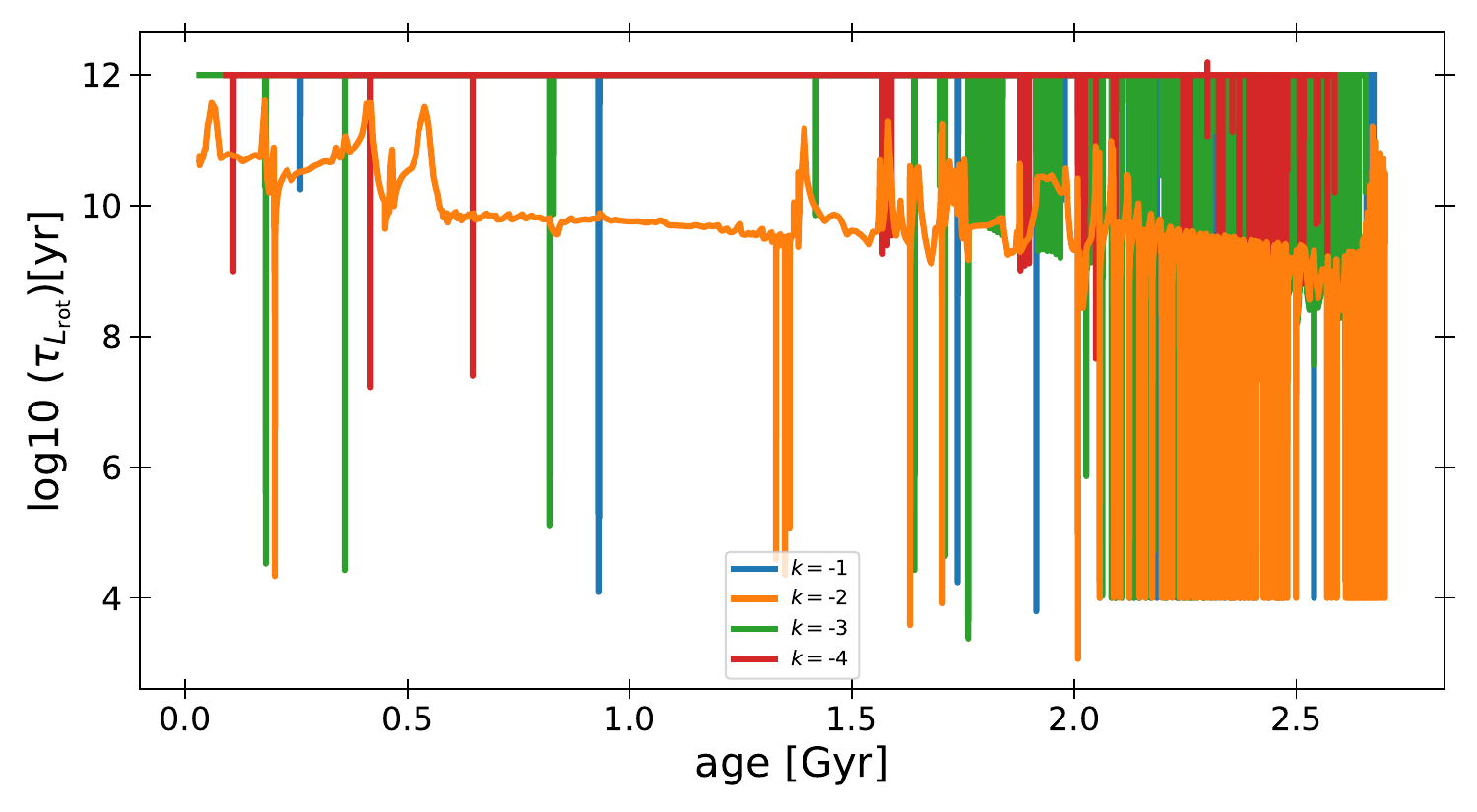}
\caption{Timescale associated with the stellar rotational angular momentum as a function of age for a twin system with low eccentricity ($e=0.1$) and high initial rotation rate ($\Omega_{\mathrm{rot}}=0.3 \Omega_{\mathrm{crit}}$), corresponding to the case shown in \autoref{fig.evolution_e0.1_hrot}. The plotted modes are retrograde, with orbital harmonics $k$ distinguished by colour. A long-term resonance locking with the $k=-2$ mode is visible as a plateau, while the peaks correspond to resonance crossings.}
\label{fig.evolution_e0.1_hrot_modes_details}
\end{figure}

In low-eccentricity systems, only a few non-negligible forcing frequencies occur, typically associated with low values of $k$. For rapidly rotating stars in wide systems, \autoref{fig.evolution_e0.1_hrot_modes_details} illustrates the detailed response of the star to the different forcing frequencies of the system.
This figure shows that for most of the main sequence, a resonance locking occurs with the $k=-2$ mode, as evidenced by the plateau in the stellar response. Outside of this locking, numerous resonance crossings are observed, corresponding to the peaks in the response. Most of these crossings occur towards the end of the main sequence, when changes in the stellar structure accelerate, leading to a faster evolution of the oscillation spectrum.

\section{Additional figures for high-eccentricty fast-rotating stars}
In the case of high-eccentricity, wide systems, many non-negligible forcing frequencies arise, which can resonate with free oscillation modes. In \autoref{fig.evolution_e0.55_hrot}, we show the detailed response of the star to a subset of these forcing frequencies.
\begin{figure}[h]
\centering
\includegraphics[width=\hsize]{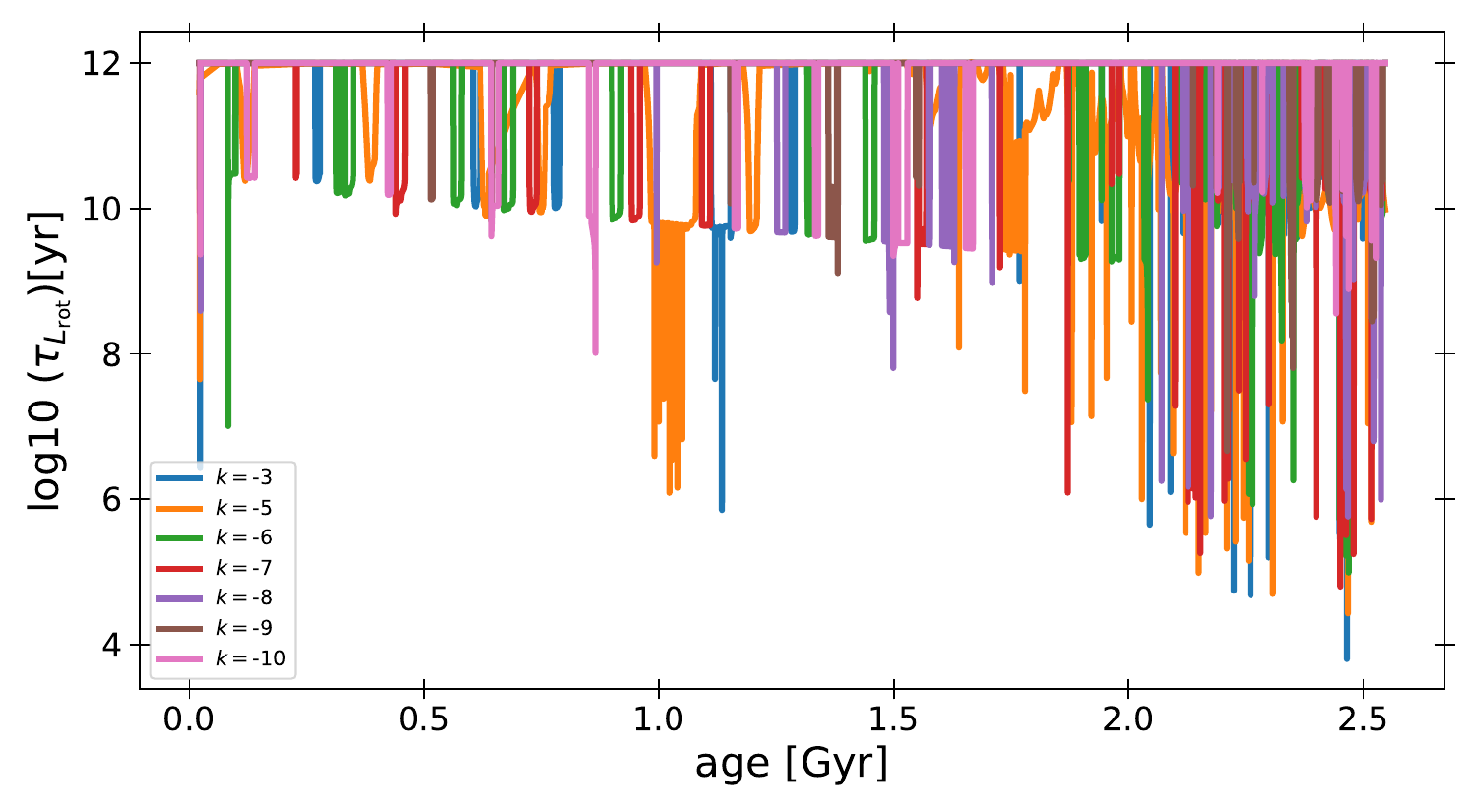}
\caption{Timescale associated with the stellar rotational angular momentum as a function of system age for a twin system with high eccentricity ($e=0.55$) and rapid initial rotation ($\Omega_{\mathrm{rot}}=0.2 \Omega_{\mathrm{crit}}$), as presented in \autoref{fig.evolution_e0.55_hrot}. The displayed modes are retrograde, with orbital harmonics $k$ indicated by colour. Resonance locking occurs successively with modes of different orbital harmonics.}
\label{fig.evolution_e0.55_hrot_modes_details}
\end{figure}
In this case, each forcing frequency is able to produce a short-lived resonance locking. The system remains locked until another, faster-moving resonance takes over, leading to a succession of brief resonance episodes. The immediate consequence is that a significant amount of angular momentum can be transferred over the course of the system’s evolution, comparable to the effect of a single long-lived locking episode.

\section{Additional figures for close-in slow-rotating stars}
\begin{figure}[h]
\centering
\includegraphics[width=\hsize]{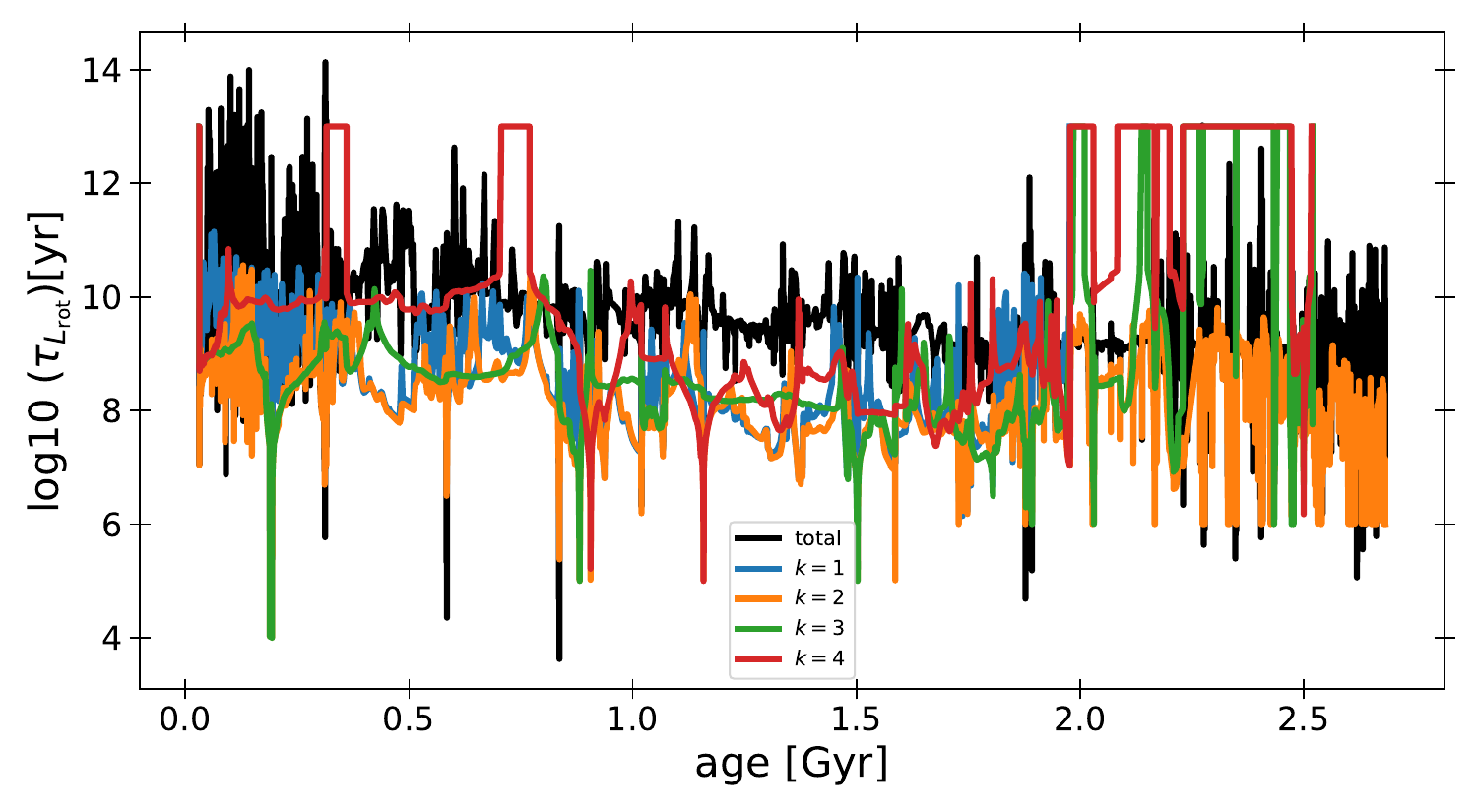}
\caption{Timescale associated with the stellar rotational angular momentum as a function of system age for a twin system with high orbital frequency and low rotation frequency, as shown in \autoref{fig.evolution_e0.1_lrot}. The plotted modes are prograde, with individual orbital harmonics $k$ shown in different colors. The black line represents the total contribution from all modes.}
\label{fig.evolution_e0.10_lrot_modes_details}
\end{figure}

In \autoref{fig.evolution_e0.10_lrot_modes_details}, we illustrate the four-mode locking occurring in a twin system with high orbital frequency and low rotation frequency, as discussed in \autoref{fig.evolution_e0.1_lrot}.
Although the amplitudes of the four prograde modes remain relatively high throughout the evolution, the total angular momentum exchange remains low. Notably, between 0.4 and 0.5 Gyr, two modes with $k = 1$ and $k = 2$ become locked simultaneously, and their contributions to the stellar angular momentum timescale nearly cancel each other out, resulting in minimal net angular momentum transfer. At other stages, all four modes participate in resonance locking, compensating each other's effects over extended timescales.

\section{Micro- and macrophysics data in MESA}\label{app:mesa}
The equation of state utilised by MESA is a blend of equations of state from the following projects, OPAL \citep{Rogers2002}, SCVH \citep{Saumon1995}, FreeEOS \citep{Irwin2004}, HELM \citep{Timmes2000}, PC \citep{Potekhin2010}, and Skye \citep{Jermyn2021}. Radiative opacities are taken primarily from OPAL \citep{Iglesias1993, Iglesias1996}, with low-temperature data from \citet{Ferguson2005} and the high-temperature, Compton-scattering dominated regime by \citet{Poutanen2017}. Electron conduction opacities are taken from \citet{Cassisi2007} and \citet{Blouin2020}. Nuclear reaction rates are from JINA REACLIB \citep{Cyburt2010}, NACRE \citep{Angulo1999} and additional tabulated weak reaction rates \citet{Fuller1985, Oda1994,Langanke2000}. Screening is included via the prescription of \citet{Chugunov2007}. Thermal neutrino loss rates are from \citet{Itoh1996}.
\section{Animation of the resonance locking}\label{apdx_animation_locking}
\begin{figure}[h]
\centering
\includegraphics[width=\hsize]{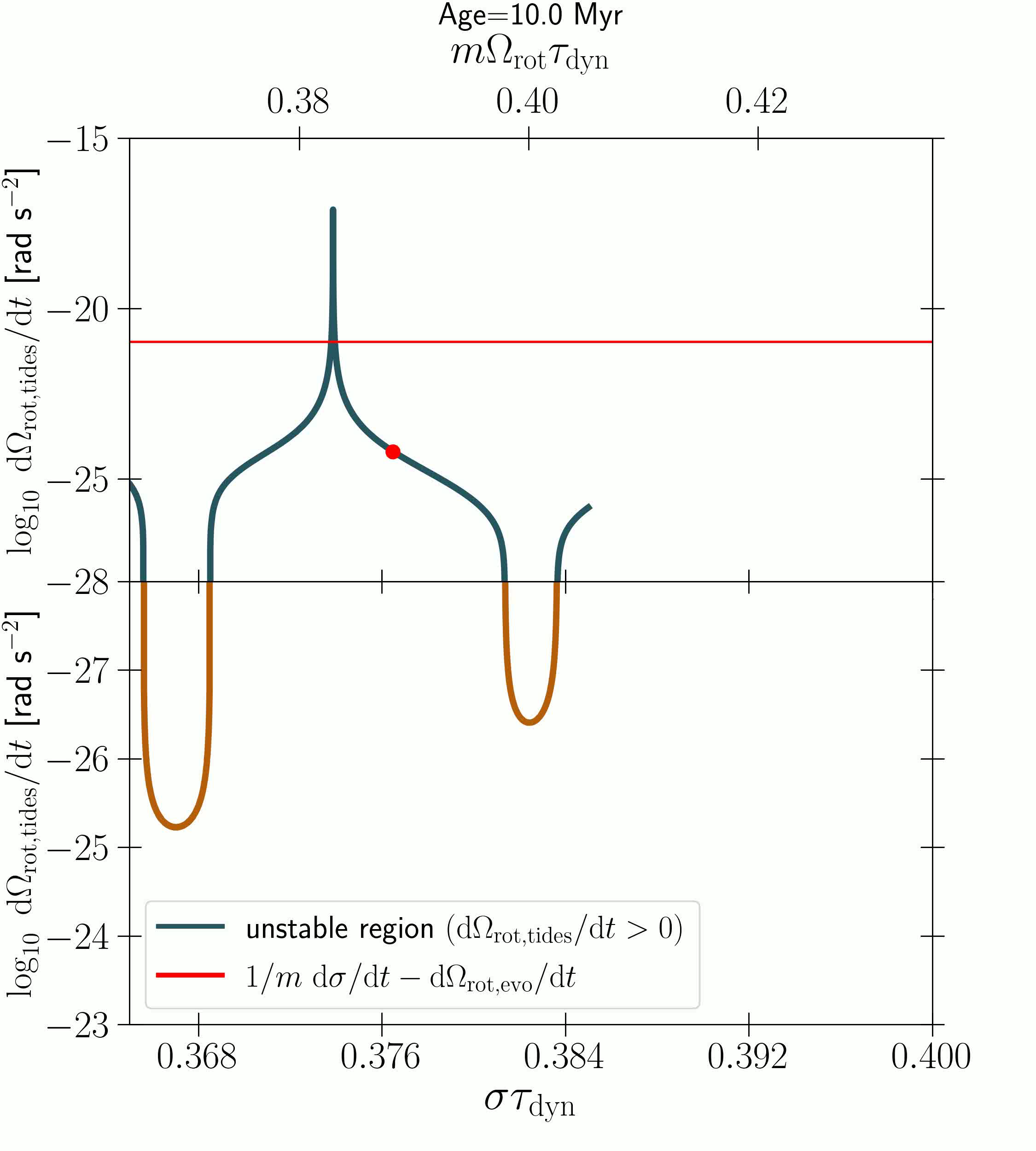}
\caption{Movie available online showing the evolution of a system forcing frequency (red dot) including the effect of dynamical tidal on the stellar rotation ($\left(\mathrm{d}\Omega_{\mathrm{rot}}/\mathrm{d}t\right)_{\mathrm{tides}}$, given on the y-axis) for a retrograde mode ($\ell=2$, $m=2$, $k=-2$, $n=30$, $\Omega{\mathrm{rot}}=0.2\  \Omega_{\mathrm{crit}}$).
The red curve gives $1/m\ \mathrm{d}\sigma/\mathrm{d}t - \left(\mathrm{d}\Omega_{\mathrm{rot}}/\mathrm{d}t\right)_{\mathrm{evo}}$, which corresponds to the possible locking positions. Blue segments represent unstable mode regions; orange curves represent stable regions. The animations starts on the right of the resonance locking position identified in \autoref{fig.illustration_resonance_locking}.}
\label{anim._full_locking}
\end{figure}

\begin{figure}[h]
\centering
\includegraphics[width=\hsize]{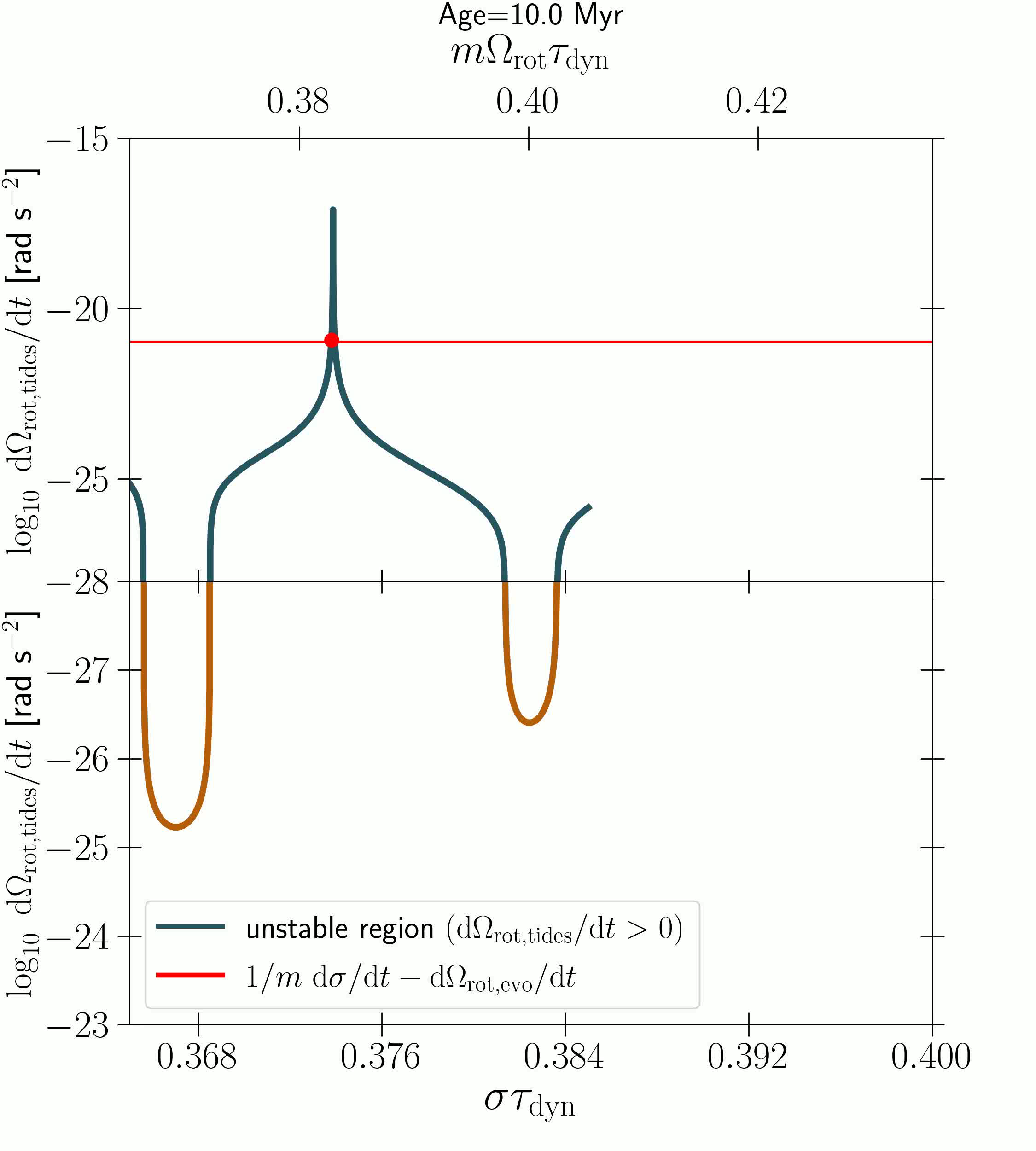}
\caption{Movie available online showing the evolution of a system forcing frequency (red dot) including the effect of dynamical tidal on the stellar rotation ($\left(\mathrm{d}\Omega_{\mathrm{rot}}/\mathrm{d}t\right)_{\mathrm{tides}}$, given on the y-axis) for a retrograde mode ($\ell=2$, $m=2$, $k=-2$, $n=30$, $\Omega{\mathrm{rot}}=0.2\  \Omega_{\mathrm{crit}}$).
The red curve gives $1/m\ \mathrm{d}\sigma/\mathrm{d}t - \left(\mathrm{d}\Omega_{\mathrm{rot}}/\mathrm{d}t\right)_{\mathrm{evo}}$, which corresponds to the possible locking positions. Blue segments represent unstable mode regions; orange curves represent stable regions. The animations starts on the left of the resonance locking position identified in \autoref{fig.illustration_resonance_locking}.}
\label{anim._jump_locking}
\end{figure}

\begin{figure}[h]
\centering
\includegraphics[width=\hsize]{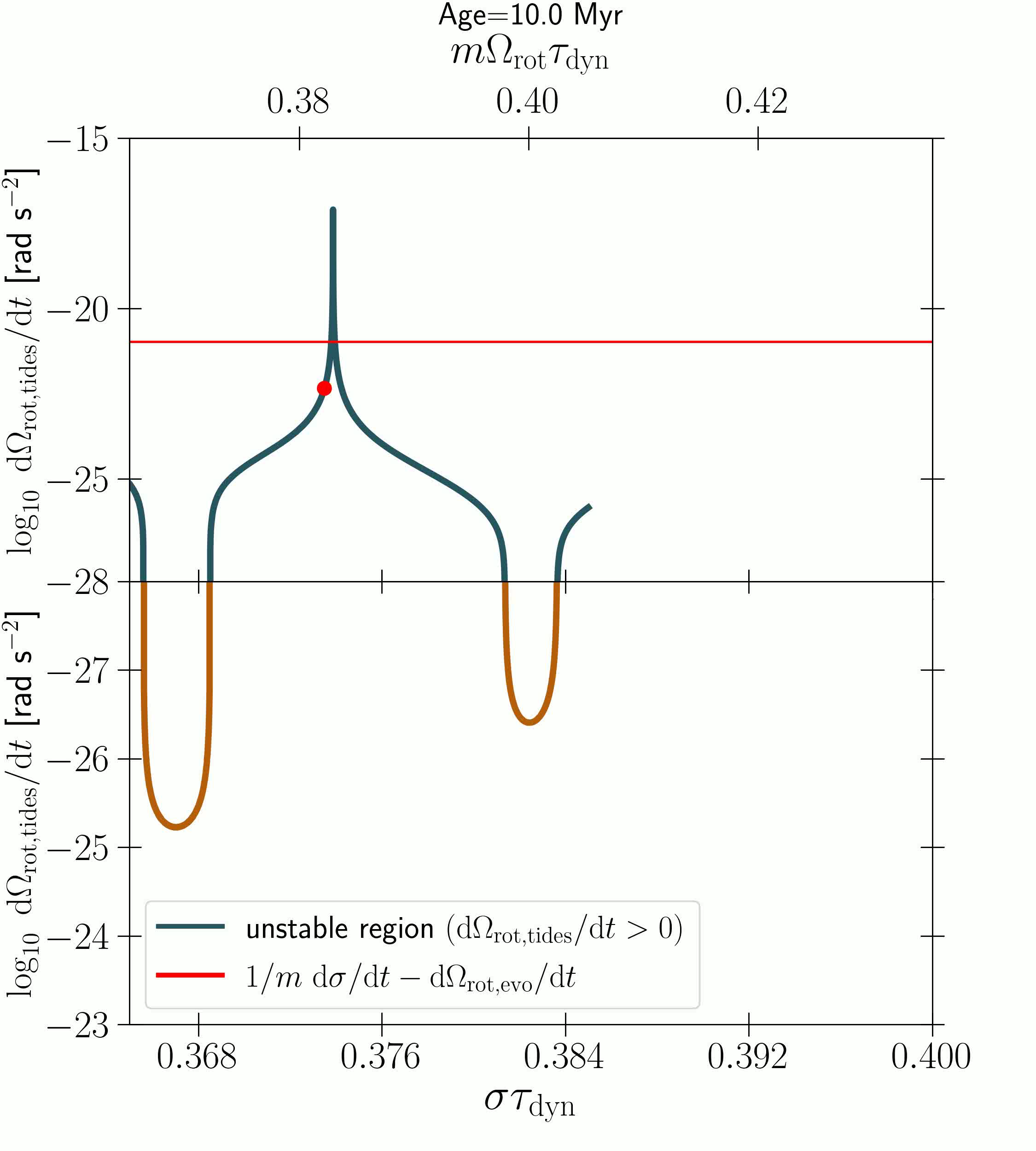}
\caption{Movie available online showing the evolution of a system forcing frequency (red dot) including the effect of dynamical tidal on the stellar rotation ($\left(\mathrm{d}\Omega_{\mathrm{rot}}/\mathrm{d}t\right)_{\mathrm{tides}}$, given on the y-axis) for a retrograde mode ($\ell=2$, $m=2$, $k=-2$, $n=30$, $\Omega{\mathrm{rot}}=0.2\  \Omega_{\mathrm{crit}}$).
The red curve gives $1/m\ \mathrm{d}\sigma/\mathrm{d}t - \left(\mathrm{d}\Omega_{\mathrm{rot}}/\mathrm{d}t\right)_{\mathrm{evo}}$, which corresponds to the possible locking positions. Blue segments represent unstable mode regions; orange curves represent stable regions. The animations starts on the far left of the resonance locking position identified in \autoref{fig.illustration_resonance_locking},  where the resonance has passed.}
\label{anim._passed_locking}
\end{figure}

\begin{figure}[h]
\centering
\includegraphics[width=\hsize]{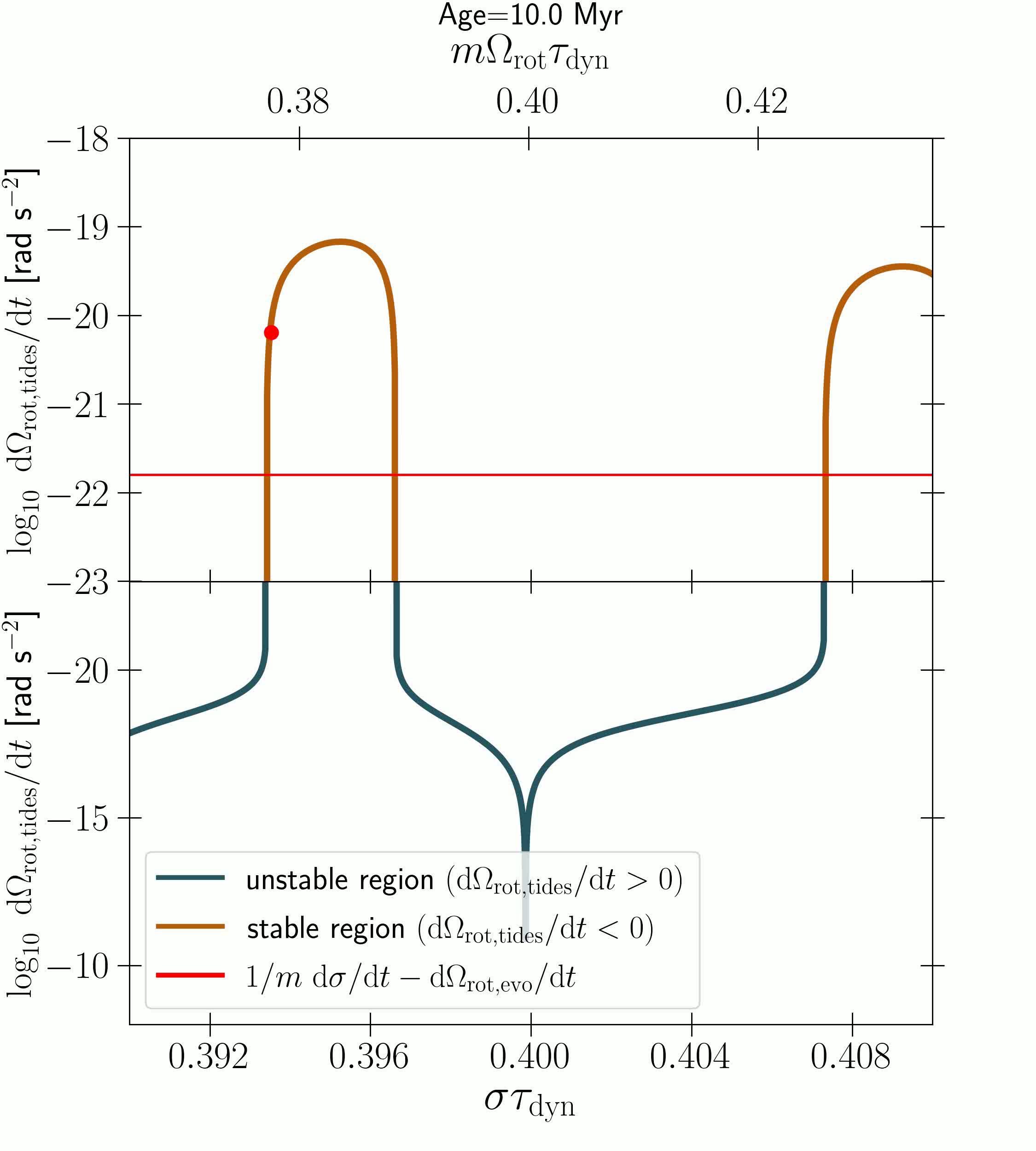}
\caption{Movie available online showing the evolution of a system forcing frequency (red dot) including the effect of dynamical tidal on the stellar rotation ($\left(\mathrm{d}\Omega_{\mathrm{rot}}/\mathrm{d}t\right)_{\mathrm{tides}}$, given on the y-axis) for a prograde mode ($\ell=2$, $m=-2$, $k=2$, $n=20$, $\Omega_{\mathrm{rot}}=0.05\ \Omega_{\mathrm{crit}}$).
The red curve gives $1/m\ \mathrm{d}\sigma/\mathrm{d}t - \left(\mathrm{d}\Omega_{\mathrm{rot}}/\mathrm{d}t\right)_{\mathrm{evo}}$, which corresponds to the possible locking positions. Blue segments represent unstable mode regions; orange curves represent stable regions. The animations starts on the left of the resonance locking position identified in \autoref{fig.illustration_resonance_locking}.}
\label{anim._jump_locking_m_2}
\end{figure}

\begin{figure}[h]
\includegraphics[width=\hsize]{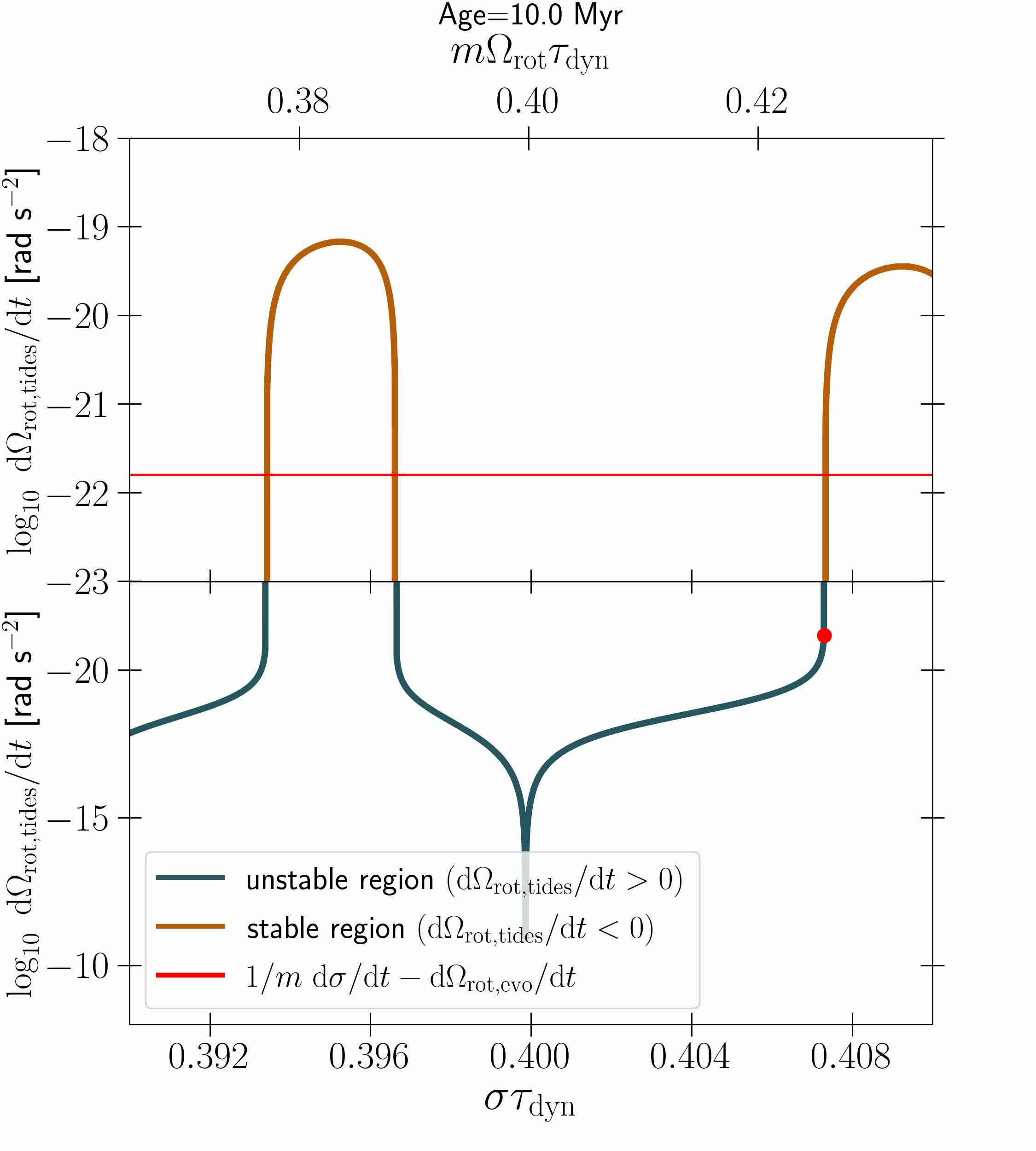}
\caption{Movie available online showing the evolution of a system forcing frequency (red dot) including the effect of dynamical tidal on the stellar rotation ($\left(\mathrm{d}\Omega_{\mathrm{rot}}/\mathrm{d}t\right)_{\mathrm{tides}}$, given on the y-axis) for a prograde mode ($\ell=2$, $m=-2$, $k=2$, $n=20$, $\Omega_{\mathrm{rot}}=0.05\ \Omega_{\mathrm{crit}}$).
The red curve gives $1/m\ \mathrm{d}\sigma/\mathrm{d}t - \left(\mathrm{d}\Omega_{\mathrm{rot}}/\mathrm{d}t\right)_{\mathrm{evo}}$, which corresponds to the possible locking positions. Blue segments represent unstable mode regions; orange curves represent stable regions. The animations starts on the far left of the resonance locking position identified in \autoref{fig.illustration_resonance_locking},  where the resonance has passed.}
\label{anim._passed_locking_m_2}
\end{figure}

In this section, we complement the explanations presented in \autoref{subsect_principle_locking} with animations illustrating the different scenarios that may occur near a resonance locking, as well as the stability of the corresponding locking positions. For these simulations, we employed a basic numerical integrator and simplified equations in order to provide a qualitative illustration of the phenomena. Consequently, these animations are only approximate representations and do not accurately describe the detailed evolution of the system. To obtain realistic results, more robust integration schemes and the full set of non-simplified equations must be used.
\\~\\In \autoref{anim._full_locking}, we show the case of a retrograde mode being drawn towards the system’s forcing frequency through stellar evolution,  inducing a long-term locking. \autoref{anim._jump_locking} illustrates the instability of the theoretical left-hand locking point shown in \autoref{fig.illustration_resonance_locking} and discussed in \autoref{subsect_principle_locking}. When the system is initialised close to this point, it rapidly evolves towards the right-hand locking position. In \autoref{anim._passed_locking}, we present the situation in which a mode has already passed a forcing frequency, causing the system to move towards the next mode and preventing a stable locking at the left-hand position identified in \autoref{fig.illustration_resonance_locking}.
\\For the prograde modes, \autoref{anim._full_locking} illustrates a mode being drawn towards a forcing frequency, leading to a stable locking. Additionally, \autoref{anim._jump_locking_m_2} demonstrates the instability of the left-hand locking point, while \autoref{anim._passed_locking_m_2} shows the case in which a retrograde mode has already passed a forcing frequency.

\begin{figure}[h]
\centering
\includegraphics[width=\hsize]{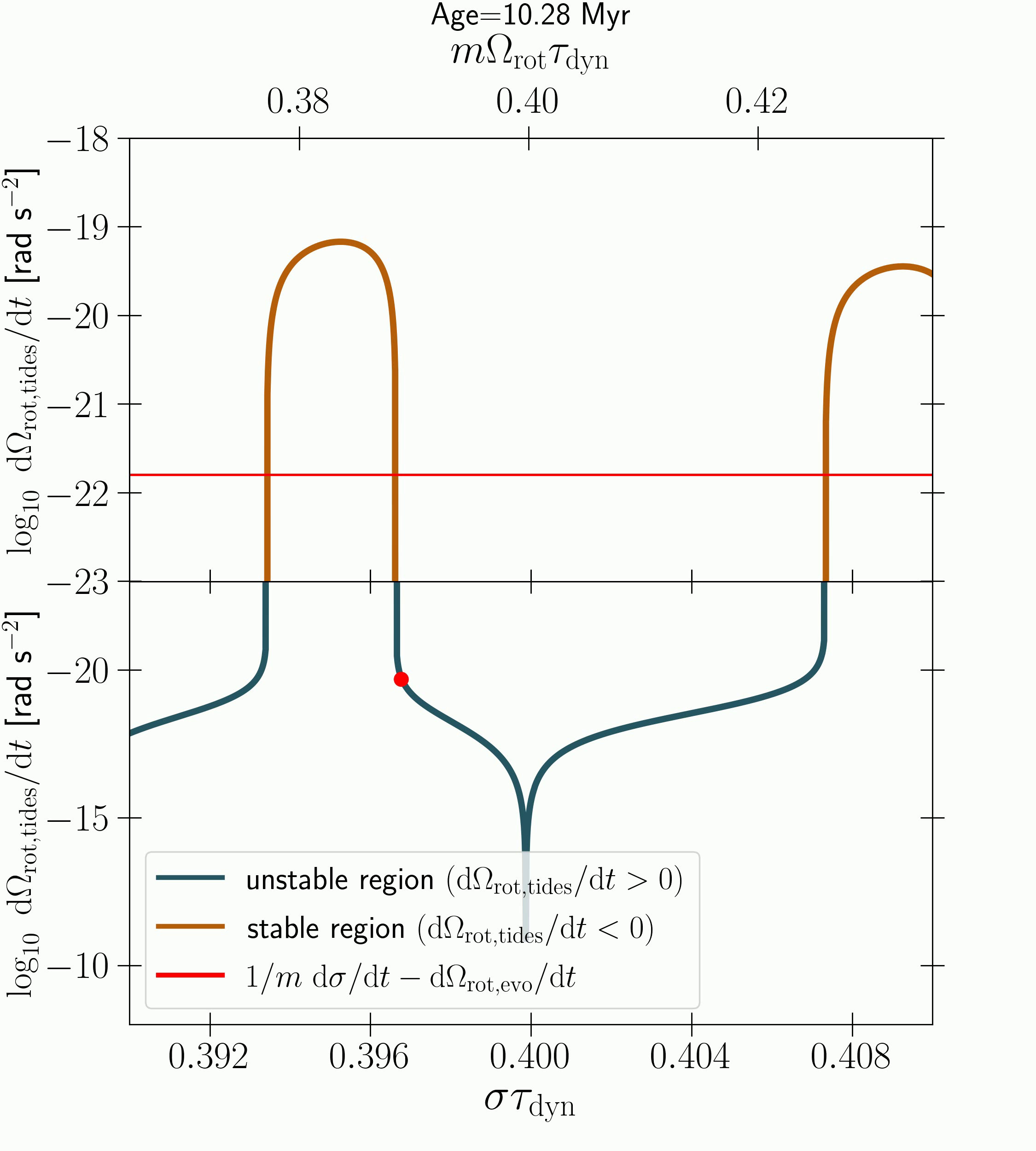}
\caption{Movie available online showing the evolution of a system forcing frequency (red dot) including the effect of dynamical tidal on the stellar rotation ($\left(\mathrm{d}\Omega_{\mathrm{rot}}/\mathrm{d}t\right)_{\mathrm{tides}}$, given on the y-axis) for a prograde mode ($\ell=2$, $m=-2$, $k=2$, $n=20$, $\Omega_{\mathrm{rot}}=0.05\ \Omega_{\mathrm{crit}}$).
The red curve gives $1/m\ \mathrm{d}\sigma/\mathrm{d}t - \left(\mathrm{d}\Omega_{\mathrm{rot}}/\mathrm{d}t\right)_{\mathrm{evo}}$, which corresponds to the possible locking positions. Blue segments represent unstable mode regions; orange curves represent stable regions. The animations starts on the right of the resonance locking position identified in \autoref{fig.illustration_resonance_locking}.}
\label{anim._full_locking_m_2}
\end{figure}

 \end{document}